\documentclass[12pt]{article}
\usepackage[top=1.2in, bottom=1.2in, left=1in, right=1in]{geometry}
\usepackage{graphicx}
\usepackage{actuarialsymbol}
\usepackage{booktabs}
\usepackage{subcaption}
\usepackage{comment}
\usepackage{tikz}
\usepackage[figuresright]{rotating}
\usepackage{tablefootnote}
\usepackage[colorlinks,
linkcolor=black,
anchorcolor=black,
citecolor=black
]{hyperref}
\usepackage{diagbox}

\usepackage[sort&compress,authoryear]{natbib}
\usepackage{BOONDOX-cal}
\usepackage{epsfig, color}
\usepackage{graphicx}
\usepackage{amsmath,amsthm,amsfonts,amssymb,bm}
\allowdisplaybreaks[3]

\newtheorem{theorem}{Theorem}[section]
\newtheorem{definition}[theorem]{Definition}

\newtheorem{proposition}[theorem]{Proposition}
\newtheorem{lemma}[theorem]{Lemma}
\newtheorem{assumption}[theorem]{Assumption}

\usepackage[figuresright]{rotating}
\usepackage{caption}
\usepackage{makeidx}
\usepackage{listings}
\title{Valuation of variable annuities under the Volterra mortality and rough Heston models}

\author{%
Wenyuan Li\thanks{Department of Statistics and Actuarial Science, The University of Hong Kong, Pok Fu Lam Road, Hong Kong, China. Email: \texttt{wylsaas@hku.hk}.}%
\qquad
Haoqi Lyu\thanks{Corresponding author. Department of Statistics and Actuarial Science, The University of Hong Kong, Pok Fu Lam Road, Hong Kong, China. Email: \texttt{lyuhq@connect.hku.hk}.}%
}
\date{\today}

\begin{document}

\maketitle
\begin{abstract}
This paper investigates the valuation of variable annuity (VA) contracts with an early surrender option under non-Markovian models. Moreover, policyholders are provided with guaranteed minimum maturity and death benefits to protect against the downside risk. Unlike the existing literature, our VA account value is linked to two non-Markovian processes: an equity index modeled by a rough Heston model and a force of mortality following a Volterra-type stochastic model. In this case, the early surrender feature introduces an optimal stopping problem where continuation values depend on the entire path history, rendering traditional numerical methods infeasible. We develop a deep signature Least Squares Monte Carlo approach to learn optimal surrender strategies on a discretized time grid. To mitigate the curse of dimensionality arising from the path-dependent model, we use truncated rough-path signatures to encode the historical paths and approximate the continuation values using a neural network. Numerically, we find that the fair fee increases with the Hurst parameters of both the stock volatility and the force of mortality. Finally, a convergence proof is provided to further support the stability of our method.  
\end{abstract}

\clearpage

\section{Introduction}
As people live longer than before, the demand for reliable and sustainable retirement income is growing. Among all actuarial products, the variable annuity (VA) is unique in meeting this demand by combining investment growth with insurance protection. Specifically, under a typical VA contract, the policyholder pays a lump sum to the insurer, who invests it in a mutual fund for profit. To mitigate downside risk, the insurer provides the policyholder with guaranteed benefits. Supported by a strong stock market performance, the VA market has experienced a resurgence after the pandemic. According to the latest LIMRA report (see \cite{LIMRA2025}),  Registered Index-Linked Annuity (RILA) sales have increased 19\% year over year to 57.4 billion US dollars in the first three quarters of 2025. Meanwhile, traditional VA sales reached 47.2 billion US dollars, 7\% higher than the same period in 2024.  

Among the VA studies, one key direction is investigating different types of guarantees. Typical examples of guarantee benefits include the Guaranteed Minimum Maturity Benefit (GMMB), which ensures a minimum payment at contract maturity, and the Guaranteed Minimum Death Benefit (GMDB), which guarantees a minimum death benefit if the insured dies before the contract matures. These guarantees are funded by fees deducted from the account value, and the manner in which fees are charged varies across products. A common choice is to set the fee rate as a constant proportional to the account value (e.g., \cite{Bernard2014IME}, \cite{Shen2016IME}, 
\cite{Jeon2018IME}, \cite{Kang2018Optimal}, and \cite{Jeon2021Pricing}). More complicated fee structures are also examined, such as VIX-linked fee (\cite{CuiFengMacKay2017VIXLinkedFees} and \cite{mackay2023analysis}), state-dependent fee (\cite{Bernard_Hardy_Mackay_2014}), and high-water mark fee structure (\cite{landriault2021high}). Moreover, many VA contracts also allow the policyholder to surrender the contract before maturity and receive a cash value linked to the current account value. The early surrender feature resembles an American-style option: at each decision time, the policyholder compares the immediate surrender payoff with the expected value of continuing to hold the contract, thereby determining whether early surrender is optimal. Consequently, valuing a VA with surrender becomes an optimal stopping problem.

The optimal stopping problem does not have an explicit solution under a finite horizon, and its solution relies heavily on numerical algorithms. In much of the existing literature, the underlying equity process is often assumed to be Markovian, so the contract value can be expressed as a function of low-dimensional state variables and analyzed by dynamic programming. There are several popular numerical algorithms for computing the optimal surrender strategy of a variable annuity. The first is the early exercise premium (EPP) method. It represents the VA value as a European part plus an early exercise premium, and the optimal surrender boundary is obtained by solving an integral equation (see \cite{Bernard2014IME}, \cite{Shen2016IME}, \cite{Jeon2018IME}, and \cite{Jeon2021Pricing}). The second is the finite difference method. It first derives the variational inequality for the VA value function, and then uses finite difference approximation to replace derivatives. The value function is finally computed iteratively using time-stepping schemes (see \cite{BrennanSchwartz1976EquityLinked}, \cite{Kang2018Optimal}, and \cite{landriault2021high}). The third is the Least Squares Monte Carlo (LSMC) approach. It uses regression to approximate continuation values and determine the optimal surrender strategy from backward computation (see \cite{BacinelloBiffisMillossovich2010RegressionBased} and \cite{shen2020pricing}). The fourth is the continuous-time Markovian chain (CTMC) approach. It adjusts the generator matrix of the CTMC to match the first and second moments of the underlying process and computes the continuation value in an exponential matrix form (see \cite{mackay2023analysis}). The fifth is the binomial-lattice (tree) method.  It uses a lattice to avoid the node explosion of a non-recombining tree model and computes the VA value through backward induction (see \cite{COSTABILE2008}). The last is the Fourier-Cosine method, which represents the continuation value of VA as a Fourier-Cosine series and determines the optimal surrender strategy backward (see \cite{kang2022valuation} and \cite{ai2024valuation}).

In our paper, we consider the valuation of the variable annuity with an early surrender feature in a non-Markovian framework. 
For the financial market, we model the equity price using rough Heston stochastic volatility model introduced by \cite{GatheralJaissonRosenbaum2018VolatilityRough} and \cite{ElEuchRosenbaum2019CharacteristicFunctionRoughHeston}. For the mortality model, we use a Volterra mortality model with a long-range dependence (LRD) feature first proposed by \cite{Wang2021Volterra}, and further extended by \cite{zhou2022stochastic} (mixed fractional Brownian motion) and \cite{jiang2024stochastic} (mixed fractional Poisson process). In contrast to the seminal work \cite{Wang2021Volterra}, we simplify the Volterra mortality model with a linear drift term, removing its deterministic part and the mean-reverting feature. This allows us to capture the mortality trend more with the LRD part rather than the deterministic part. Since the Volterra mortality model is non-Markovian, it is impractical to solve it through standard methods from the VA literature. To overcome this difficulty, we propose a deep signature Least Squares Monte Carlo (LSMC) approach that extends the traditional LSMC method (\cite{LongstaffSchwartz2001}) to non-Markovian markets. Specifically, it uses the path signature to encode the path information and applies deep regression to compute the VA's continuation value.  Finally, the optimal surrender strategy is determined by comparing the immediate surrender payoff and the continuation value, and the fair fee is obtained through a bisection method. 

The key challenge in extending the LSMC method to the path-dependent setting is that the continuation value depends on the entire path history. Due to the high-dimensional state variables (entire path), the traditional polynomial regression method becomes infeasible. Therefore, we consider a path-signature encoding to reduce the dimensionality of the state variable. Path signatures originate from rough path theory. \cite{lyons1998differential} first introduces the use of iterated integrals (signatures) to represent paths and proposes a framework to solve differential equations driven by rough signals. After that,  signature-based methods are  widely applied in option pricing under the path-dependent settings, such as \cite{Bayraktar2024} and \cite{jaber2025signature} for options with path-dependent payoffs, and 
\cite{bayer2025pricing}, \cite{Shah2025AmericanOptionPricingTimeVaryingRoughVolatility}, and \cite{ma2025option} for path-dependent volatility models.

Beyond the numerical algorithm, we also provide a convergence analysis for the deep signature LSMC scheme used to solve the optimal surrender problem for variable annuities. The proof decomposes the pricing error into (i) Monte Carlo sampling error, (ii) time discretization error, (iii) signature truncation error, and (iv) neural network approximation error. We show that these error components vanish as numerical resolutions increase. These results provide theoretical support to ensure the stability of the numerical procedure.

 Our numerical experiments analyze the effect of rough Heston market and Volterra mortality on the insurer's fair fee and the policyholder's surrender strategies. In particular, we investigate the sensitivity of the fair fee to the roughness of stock volatility and the long-range dependence in mortality intensity. We find that the fair fee monotonically increases with the Hurst parameters of both the rough volatility stock model and the Volterra mortality model. We also provide a visualization of the optimal surrender behavior. Since the continuation value depends on the full history under rough volatility and Volterra mortality, there is no low-dimensional free-boundary for the surrender rule. We therefore plot a three-dimensional projection of exercise/continuation decisions across time, fund value, and volatility, show representative sample paths labeled by the learned surrender time, and report the surrender rates over time. These figures offer an intuitive explanation of the optimal stopping rule in the path-dependence setting.

Our paper contributes to the variable annuity literature by incorporating path-dependent equity and mortality models. Using the deep signature LSMC algorithm, we can solve the non-Markovian optimal stopping problem numerically, even though most traditional methods fail. Our contributions are summarized below:
\begin{itemize}
    \item We develop a valuation framework for variable annuities with an early surrender option under a joint non-Markovian model that combines rough volatility and Volterra-type mortality.
    \item We propose a numerical algorithm using LSMC with signature encoding and deep regression to approximate the continuation value of the VA contract, and solve the fair fee numerically by the bisection method.
    \item We decompose the computation error into Monte Carlo sampling error, signature truncation error, neural network approximation error, and time discretization error, and prove the convergence of the deep signature LSMC algorithm.
    \item We provide a comprehensive numerical study of the fair fee rates and surrender behavior. We illustrate the impact of the stock volatility roughness and long-range dependence of the Volterra mortality on VA pricing and surrender strategies.
\end{itemize}

The rest of the paper is organized as follows. Section~\ref{sec: problem_statement} specifies the Volterra mortality model, the rough Heston market dynamics, and the VA contract features. Section~\ref{sec: Optimal_surrender} presents the formulation of the optimal surrender problem, illustrates the deep signature LSMC algorithm, and specifies the neural network architecture. Section~\ref{sec: convergence} analyzes the convergence of the algorithm. Section~\ref{sec: numerical} shows numerical examples, including the sensitivity analysis of model parameters and visualizations of surrender strategies. Section~\ref{sec: conclusions} briefly concludes the paper.

\section{Problem statement}
\label{sec: problem_statement}
We consider a variable annuity (VA) contract subject to both mortality risk with long-range dependence (LRD) and financial market risk with path dependence. Let us consider a filtered probability space $(\Omega, \mathcal{F}, \{\mathcal{F}_t\}, \mathbb{P})$ satisfying the usual conditions. The filtration $\{\mathcal{F}_t\}_{t\in[0,T]} = \{\mathcal{F}_t:0\leq t \leq T\}$ is the joint filtration generated by the Brownian motions driving the financial market and mortality models. When considering the GMDB in VA pricing, we cannot hedge mortality risk by trading stocks and bonds, which creates an incomplete market. To overcome this difficulty, the actuarial literature usually makes the following two common assumptions: independence between financial and mortality risks, and the insurer's risk-neutrality with respect to mortality risk. 
Therefore, we can define the risk-neutral measure for VA pricing as the product measure combining the financial market measure $\mathbb{Q}$ and the physical measure $\mathbb{P}$ for mortality risk. To avoid abusing the notation, we denote this product measure by $\mathbb{Q}$. See the same arguments in \cite{aase1994pricing}, \cite{bauer2008universal}, and \cite{landriault2021high}.

\subsection{Mortality model}
\label{sec: mortality_model}
We assume the mortality follows a simplified form of the Volterra mortality model proposed by \cite{Wang2021Volterra}. Suppose a policyholder's initial age is $x$, and their force of mortality after $t$ years is denoted by \( \mu_{x+t} \), which follows a stochastic Volterra integral equation (SVIE)
\begin{equation}
    \mu_{x+t} = \mu_x + \int_0^t K_m(t-u) b(\mu_{x+u}) du + \int_0^t K_m(t-u) \sigma(\mu_{x+u}) d\mathbf{B}_u,
    \label{eq: stochastic_vol}
\end{equation}
where \( K_m(\cdot) \) is a kernel function for the mortality model and \( \mathbf{B} = [B_1, \dots,B_d]\) is a standard $d-$dimensional Brownian motion under physical measure $\mathbb{P}$. The covariance matrix is defined as
\[
a(\mu):=\sigma(\mu)^{\top}\sigma(\mu),
\]
where $a(\mu)$ and $b(\mu)$ follow affine forms
\begin{align*}
    a(\mu) &= A^0 +(\mu^1)^{\top}A^1+\cdots+(\mu^d)^{\top}A^d, \\
    b(\mu) &= b^0 +(\mu^1)^{\top}b^1+\cdots+(\mu^d)^{\top}b^d,
\end{align*}
$A^i$ are $d-$dimensional symmetric matrices and $b^i$ are column vectors, and the quadratic term of $A^i$ can be defined as $A(v):=(v^{\top}A^1v, \dots , v^{\top}A^dv)$ for some $d-$dimensional column vector $v$. We set $B=(b^1,\cdots,b^d)$.

Let $T_{x+t}$ denote the future lifetime of an insured aged $x+t$. The survival probability is denoted by 
\[
\px[s-t]{x+t}:=\mathbb{P}(T_{x+t}>s-t|\mathcal{F}_t),
\]
which is the probability that a person will survive from time $t$ to time $s$ (from age $x+t$ to age $x+s$). Corollary 1 in \cite{Wang2021Volterra} proves that the survival probability can be calculated by 
\begin{equation}
    \px[s]{x} = \mathbb{P}(T_{x}>s)=\mathbb{E}\left[e^{-\int_0^s\mu_{x+u}du}\right]= \exp(Y_0(s)),
    \label{eq: survival_prob}
\end{equation}
where
\begin{equation}
    Y_0(s)=\int_0^s(-\eta \mu_x+\psi(u)^{\top}b(\mu_x)+\dfrac{1}{2}\psi(u)^{\top}a(\mu_x)\psi(u))du,
    \label{eq: Y0}
\end{equation}
and $\psi \in \mathcal{L}^2([0, T], \mathbb{C}^d)$ solves the Riccati–Volterra equation
\begin{equation}\label{eq:psi}
\psi = \left( -\eta + \psi^{\top} B + \frac{1}{2} A(\psi) \right) * K_m.    
\end{equation}

For the VA valuation problem, we adopt a one-factor model that preserves the long-range dependence induced by the Volterra integration. We set the dimension $d=1$ and choose $a(\cdot),b(\cdot)$ such that $\mu_{x+t}$ follows an affine Volterra process. We parameterize the SVIE as
\begin{equation}
    \mu_{x+t} = \mu_x + \lambda\int_0^tK_m(t-u) \mu_{x+u}du+\int_0^tK_m(t-u)\sigma\sqrt{\mu_{x+u}}dB_u,
    \label{eq: SVIE_mort}
\end{equation}
where $K_m(t):=t^{H_m-\frac{1}{2}}/\Gamma(H_m+\frac{1}{2})$ is the fractional kernel for mortality, $H_m \in (0,1)$ is the Hurst parameter for mortality, and subscript ``$m$'' is short for ``mortality''. When $H_m=1/2$, the SVIE \eqref{eq: SVIE_mort} degenerates to a Markovian process driven by Brownian motions. When $H_m\in(0,1/2)$, SVIE \eqref{eq: SVIE_mort} has a rough path with short-term memory. When $H_m\in(1/2,1)$, SVIE \eqref{eq: SVIE_mort} has a smooth path with long-range dependence (LRD). We will use mortality data to determine the value of $H_m$. In the notation of (\ref{eq: stochastic_vol}), our choices correspond to
\[
b(\mu)=\lambda\mu,a(\mu)=\sigma^2\mu.
\]

In the seminal work of \cite{Wang2021Volterra}, the mortality process includes a mean-reversion drift term (i.e., $b(\mu)=\lambda(\theta-\mu), \text{ }\lambda, \theta>0$) and an additional deterministic component $m(t)$. From their numerical results, we observe that the deterministic component explains more of the information, and the non-Markovian stochastic part explains little. To demonstrate the effectiveness of the Volterra mortality model, we make the following modifications: remove the deterministic part entirely and replace the mean-reverting drift with a linear drift. Notably, it remains controversial whether the force of mortality exhibits a mean-reverting pattern (see \cite{luciano2005non} and \cite{shen2018lifetime}). In this study, we calibrate the model with the U.S. data from the Human Mortality Database and find that a linear drift term outperforms others. The reason is that excluding the mean-reverting feature in the drift term allows the model to better capture the long-run upward trend of the force of mortality.

From equation (\ref{eq: survival_prob}), the probability that an individual aged $x$ can survive to time $s$ is given by
\[
\px[s]{x}(H_m, \lambda, \sigma, \mu_x) = \exp(Y_0(s)).
\]
The survival probability above is a function of model parameters $(H_m, \lambda, \sigma, \mu_x).$ In calibration, we fix the initial mortality intensity as $\mu_x=-\ln(l_{x+1}/\l_x)$ (see \cite{zhou2022stochastic} and \cite{jiang2024stochastic}) and estimate $(H_m, \lambda, \sigma)$ by fitting model-implied survival probabilities to those observed in the life table. We calibrate the mortality model using US males born in 1920, aged 30 to 80, with data from the Human Mortality Database (HMD). Inspired by the early work \cite{zhou2022stochastic} and \cite{jiang2024stochastic}, we estimate the model parameters by minimizing the mean square error (MSE) between the model-implied and observed survival probabilities for individuals aged $x$
\[
H_m^*, \lambda^*, \sigma^* =\arg \min_{H_m, \lambda, \sigma} \dfrac{1}{N}\sum_{s=1}^N\left(\px[s]{x}(H_m, \lambda, \sigma, \mu_x)-\dfrac{l_{x+s}}{l_x}\right)^2,
\]
where $l_x$ denotes the number of survivors of US males of cohort 1920 obtained from the life table, the initial mortality intensity is set to $\mu_x = -\ln (l_{x+1}/l_x)$. The estimated parameters are reported in Table \ref{table:params_mort_cali}. $H_m=0.703923>1/2$ implies that the force of mortality has a smooth path with long-range dependence.

\begin{table}[h!]
\centering
\caption{Values of Estimated Parameters }
\begin{tabular}{cccc}
\hline
$H_m$ & $\lambda$ & $\sigma$ & $\mu_x$ \\
\hline
$0.703932$ & $0.047780$ & $0.005023$ & $0.002102$ \\
\hline
\end{tabular}
\label{table:params_mort_cali}
\end{table}

\begin{figure}[h!]
        \centering
    \includegraphics[width=0.6\linewidth]{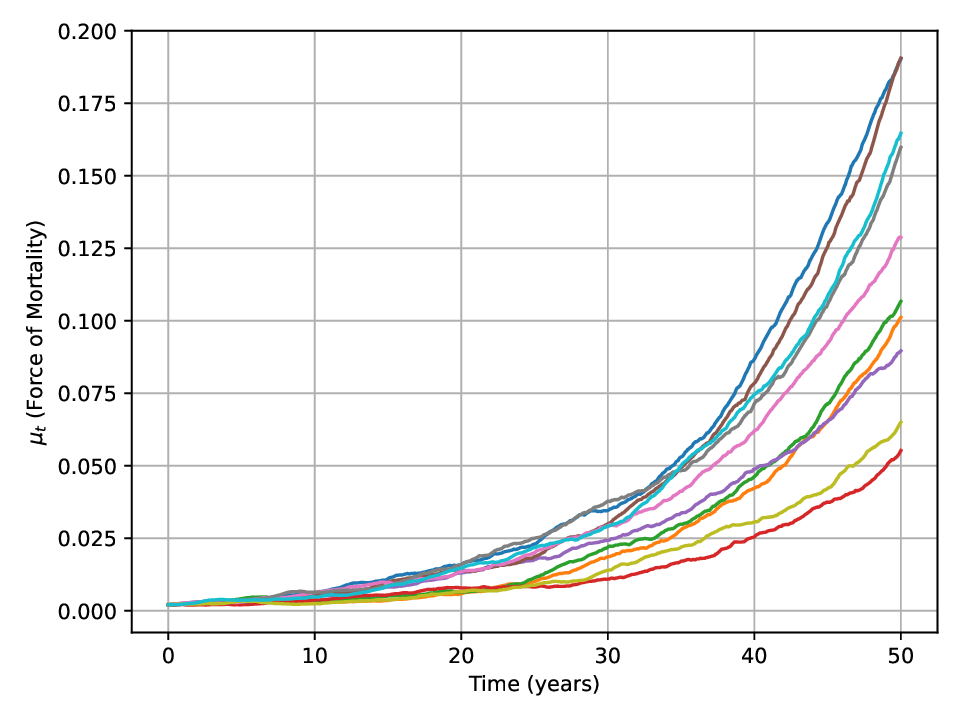}
        \caption{Sample Paths of the Stochastic force of Mortality}
        \label{fig:sample_mortality_30_80}
\end{figure}

\begin{figure}[h!]
        \centering
    \includegraphics[width=0.6\linewidth]{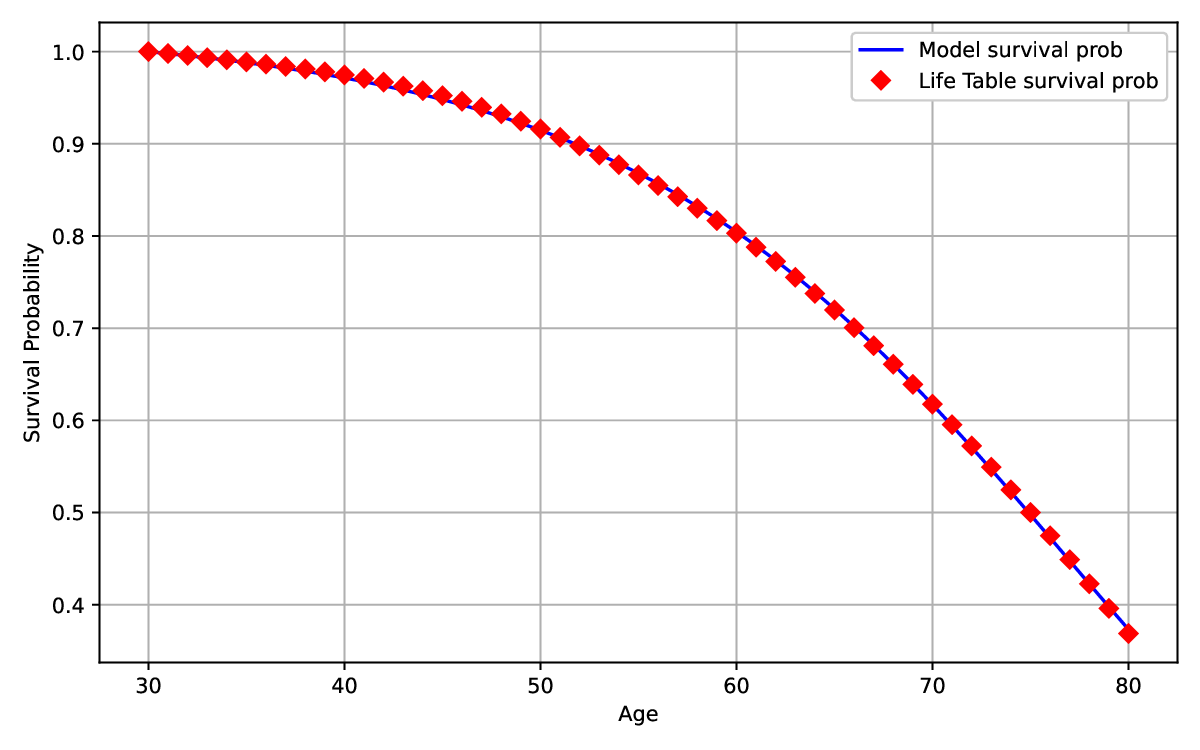}
        \caption{Observed survival probabilities and model survival
        probabilities}
        \label{fig:survival_30_80}
\end{figure}

 Furthermore, Figure \ref{fig:sample_mortality_30_80} shows sample paths of the stochastic force of mortality generated under the calibrated parameters, illustrating the variability and the long-term upward trend induced by the stochastic model. Figure \ref{fig:survival_30_80} compares the model-implied survival probability curve with the empirical survival probabilities from the cohort life table. The close agreement indicates that the model fits the survival probabilities very well.

\subsection{Market model and variable annuity account}
We consider a VA policy with maturity $T$. The initial account value $F_0$ equals the initial premium paid by the policyholder. The VA account tracks an underlying equity index with price process $S=\{S_t:0\le t\le T\}$. Under the risk-neutral measure \( \mathbb{Q} \), we assume that the equity value process $S_t$ follows a rough Heston model and is independent of the mortality process. The dynamics of the equity value are given by

\begin{equation}
\begin{cases}
    dS_t = rS_tdt+S_t \sqrt{V_t} dW_t, & S_0 = s_0, \\
    V_t = V_0 + \int_0^t K_S(t-s) \Big[ \gamma(\theta - V_s) ds + \nu \sqrt{V_s} dZ_s \Big], & V_0 = v_0, \\
    d\langle W, Z \rangle_t = \rho dt,
\end{cases}
\label{eq: heston}
\end{equation}
where \( K_S(t) := t^{H_S - \frac{1}{2}}/\Gamma(H_S+\frac{1}{2}) \) is the kernel function, $H_S$ is the Hurst parameter for the stock price model, the subscript ``$S$'' is short for ``stock'', \( r \) is the constant risk-free rate, $ \gamma, \theta, \nu$ are model parameters, and \( W \), \( Z \) are standard Brownian motions with correlation \( \rho \).

Let \( F = \{ F_t : 0 \leq t \leq T \} \) denote the value of the variable annuity  account at time \( t \). A management fee is deducted continuously at a constant rate $c\ge0$, so that
\begin{equation}
F_t =S_t e^{-ct}.
\end{equation}

In our numerical study, the rough Heston parameters $(H_S, \gamma, \theta, \nu, \rho, V_0)$ are taken from \cite{Jeng2021SPX}, calibrated to equity index option data. Parameter values are summarized in Section~\ref{sec: numerical}.

\subsection{Guaranteed benefits and early surrender}

We consider a VA contract offering two types of guaranteed benefits: the Guaranteed Minimum Maturity Benefit (GMMB) and the Guaranteed Minimum Death Benefit (GMDB). Specifically, the maturity payoff is $\max(G, F_T)$, where $G$ is a constant guaranteed benefit, if the policyholder survives to maturity $T$ and the contract has not been surrendered. Furthermore, if the policyholder dies at time $t<T$ before surrender, then the death benefit paid at $t$ is $\max(G, F_t)$. 

The policyholder may surrender the VA at any time prior to maturity. If the contract terminates at time $t\le T$ due to surrender ($t<T$) or maturity ($t=T$), the policyholder receives
\begin{equation}
g_t:=
\begin{cases}
(1-\kappa_t)F_t, & 0\le t<T, \\
\max(G, F_T), & t=T,
\end{cases}
\label{eq: payoff_g}
\end{equation}
where $\kappa_t$ is a surrender penalty charged by the insurer to discourage early surrender. Following \cite{Bernard2014IME}, \cite{Shen2016IME}, and \cite{Jeon2018IME}, 
we set 
\[
\kappa_t=1-e^{-\kappa(T-t)},
\]
with constant $\kappa>0$, so that the surrender value equals $e^{-\kappa(T-t)}F_t$ and the penalty decreases as maturity approaches.

\section{Fair fee and optimal surrender strategy}
\label{sec: Optimal_surrender}

This section presents the fair fee definition and our numerical procedure for computing it under the optimal surrender decision. The fair fee $c^*$ is the fee rate such that the time-zero VA value equals the initial premium. For a given fee level $c$, the corresponding time-0 VA value $U_0(c)$ is obtained by solving a non-Markovian optimal stopping problem via a deep signature LSMC approach on a discretized grid (Sections \ref{sec: Forward simulation of the underlying process}-\ref{sec: Deep signature LSMC method}). We then solve $U_0(c)=F_0$ for $c^*$ using the bisection method (Section \ref{sec: Fair fee}).

\subsection{Forward simulation of the underlying process}
\label{sec: Forward simulation of the underlying process}
Implementing the deep signature LSMC algorithm requires generating Monte Carlo sample paths of the underlying stochastic process from time $0$ to time $T$. We partition the continuous time interval $[0,T]$ by a discretized grid $0 = t_0 < t_1 < \cdots < t_N = T$ and the uniform step size is given by $h := t_{n+1} - t_n$.

For the stock process, the variance of the rough Heston model is dependent on the entire historical path through kernel $(t-s)^{H_S-\frac{1}{2}}$. Therefore, a direct Euler scheme costs $O(N^2)$ per path, with $N$ being the discretization steps. To reduce computation cost, we apply the ``fast algorithm" (\cite{Ma2022Fast}, Algorithm 2) to sample rough Heston paths by approximating the kernel with a sum of exponentials, so that we can update the history recursively and reduce the cost to $O(N\log N)$. The key idea of the algorithm is to write the kernel into an integral form and approximate the integration by 
\begin{equation}\label{eq:kernel_approx}
\begin{aligned}
t^{H_S-\frac{1}{2}}
&=\frac{1}{\Gamma(\frac{1}{2}-H_S)}\int_0^\infty e^{-ts}s^{-H_S-\frac{1}{2}} \, ds\\
&=\frac{1}{\Gamma(\frac{1}{2}-H_S)}\left(
\int_0^{2^{-M^\prime}} e^{-ts}s^{-H_S-\frac{1}{2}} ds
+\sum_{j=-M^\prime}^{-1}\int_{2^j}^{2^{j+1}} e^{-ts}s^{-H_S-\frac{1}{2}} ds \right.\\
&\qquad\left.
+\sum_{j=0}^N \int_{2^j}^{2^{j+1}} e^{-ts}s^{-H_S-\frac{1}{2}} ds
\right).
\end{aligned}
\end{equation}

The integrals in \eqref{eq:kernel_approx} can be approximated by a weighted sum of exponential functions
\[
t^{H_S-\frac{1}{2}}\approx \sum_{l=1}^{N_{\text{exp}}}\omega_l e^{-x_lt},
\]
where
\begin{align*}
    N_{\text{exp}}=& \left\lfloor O\left(\log\frac{1}{\xi}(\log\log\frac{1}{\xi}+\log\frac{T}{h})+\log\frac{1}{h}(\log\log\frac{1}{\xi}\log\frac{1}{h})\right)\right\rfloor,
\end{align*}
where $x_l \text{ and } \omega_l \text{ for } l=1,\cdots, N_{\exp}$ are the union of nodes and weights determined by Gauss-Jacobi quadrature and Gauss-Legendre quadrature, and $\xi$ is a constant representing the absolute error tolerance for the approximation of the kernel function. See \cite{Ma2022Fast} for details.

Consequently, the variance process is simulated by 
\begin{align*}
    V_{t_n}=&V_0+\dfrac{h^{H_S+\frac{1}{2}}}{\Gamma(H_S+\frac{3}{2})}\gamma(\theta-V_{t_{n-1}})+\dfrac{1}{\Gamma(H_S+\frac{1}{2})}\sum_{l=1}^{N_\text{exp}}\omega_le^{-x_lh}H_l(t_{n-1})+\dfrac{h^{H_S}}{\Gamma(H_S+\frac{1}{2})}\nu Z_{t_n}\\
    &+\dfrac{1}{\Gamma(H_S+\frac{1}{2})}\sum_{l=1}^{N_{\text{exp}}}\omega_le^{-x_lh}J_l(t_{n-1}),
\end{align*}
where
\begin{align*}
    H_l(t_{n-1})=&\dfrac{\gamma(\theta-V_{t_{n-2}})}{x_l}(1-e^{-x_lh})+e^{-x_lh}H_l(t_{n-2}),\\
    J_l(t_{n-1})=&e^{-x_lh}\nu\sqrt{h}Z_{t_{n-1}}+e^{-x_lh}J_l(t_{n-2}).
\end{align*}


The kernel function of the Volterra mortality model, $t^{H_m-1/2}$, cannot be written in the integral form as in \eqref{eq:kernel_approx} because $H_m\in(0.5,1)$ and the integral
$\int_0^\infty e^{-ts}s^{-H_m-1/2}ds$ does not converge. Therefore, the ``fast algorithm" is not directly applicable to the mortality model, and we apply the standard Euler scheme for sampling the force of mortality paths. The Ito integrals are approximated by the sum of left points, and the ordinary Riemann-Stieltjes integrals with respect to time are computed using the standard trapezoidal rule. Specifically, for equations \eqref{eq: Y0}-\eqref{eq: SVIE_mort}, we have
\begin{align*}
\psi(t_n) 
=& \sum_{k=0}^{n-1} \frac{h}{2} \left( \frac{(t_n - t_k)^{H_m+\frac{1}{2}}}{\Gamma(H_m+\frac{1}{2})}\left(-\eta - \lambda\psi(t_k)\right) + \frac{(t_n - t_{k+1})^{H_m-\frac{1}{2}}}{\Gamma(H_m+\frac{1}{2})}\left(-\eta - \lambda\psi(t_{k+1})\right) \right),\\
Y_0(t_n) =& \sum_{k=0}^{n-1} \frac{h}{2} \left( 
    -\eta X_0 + \psi(t_k)\lambda X_0
    -\eta X_0 + \psi(t_{k+1})\lambda X_0
\right),\\
 \mu_{x+t_n} =& \  \mu_x 
 + \lambda \sum_{k=0}^{n-1} \frac{h}{2}\left( 
\frac{(t_n - t_k)^{H_m-\frac{1}{2}}}{\Gamma(H_m+\frac{1}{2})} \mu_{x+t_k}
+ \frac{(t_n - t_{k+1})^{H_m-\frac{1}{2}}}{\Gamma(H_m+\frac{1}{2})} \mu_{x+t_{k+1}}
\right) \\
&+ \sum_{k=0}^{n-1}  
\frac{(t_n - t_k)^{H_m-\frac{1}{2}}}{\Gamma(H_m+\frac{1}{2})}
\sigma\Delta B_k,
\end{align*}
where $\Delta B_k=B_{k+1}-B_k$.

\subsection{Dynamic programming principle for variable annuity}
\label{sec: DPP_for_VA}

The early surrender feature turns the valuation of the VA into an optimal stopping problem. At each time $t \in [0, T]$, the policyholder chooses between surrendering immediately and continuing the contract. Let $\mathcal{T}_{[t,T]}$ denote all the stopping times $\tau$ valued in $[t,T]$. Then, the VA value process $\{U_t\}_{t\in[0, T]}$ is defined as the maximal expected present value of the future cash flows over all admissible stopping times, i.e.
\begin{equation}
U_t:=\sup_{\tau \in \mathcal{T}_{[t, T]}}\mathbb{E}^{\mathbb{Q}}\left[ \px[\tau-t]{x+t} g_\tau e^{-r(\tau-t) } + \int_t^\tau \px[s-t]{x+t} \mu_{x+s} \max(G, F_s)e^{-r(s-t)}ds\bigg| \mathcal{F}_t\right],
\label{eq: Ut}
\end{equation}
where $g_t$ is the payoff function defined in \eqref{eq: payoff_g} and $\mathbb{Q}$ is a pricing measure defined at the beginning of the Section
\ref{sec: problem_statement}. The first term in \eqref{eq: Ut} represents the present value of the possible surrender benefit (if the policyholder surrenders prior to death or maturity, i.e.,$\tau<T$) or maturity benefit (the policyholder survives to the maturity time and does not surrender the VA early, i.e., $\tau=T$). The second term is the present value of the death benefit (if death occurs before maturity). In particular, the time-zero price is $U_0$ (take $t=0$). To calculate $U_0$, we propose a deep signature LSMC algorithm to approximate the optimal decision rule by a simulation and regression-based method.

We apply the dynamic programming principle (DPP) to link the value function at the current time $t$ to its value at the next decision time $t + h$.  The discretized optimal stopping problem gives the recursive relationship
\begin{align*}
U_{t_N} &= \max(G, F_{t_N}),\notag\\
    U_{t_n} &= \max\left\{ g_{t_n}, \;\;
    \mathbb{E}^{\mathbb{Q}} \left[ e^{-rh} \, \px[h]{x+t_n} \, U_{t_{n+1}} 
    + e^{-rh} (1-\px[h]{x+t_n})  \max(G, F_{t_{n+1}}) \mid \mathcal{F}_{t_n} \right]
    \right\}.
\end{align*}
    For simplicity, we denote by $A_n$ the event that the surrender payoff is higher than the expected continuation payoff at time $t_n$
    \begin{equation}
         A_n:=\left\{g_{t_n}\geq\mathbb{E}^{\mathbb{Q}}\left[ e^{-rh} \, \px[h]{x+t_n} \, U_{t_{n+1}} 
    + e^{-rh} (1-\px[h]{x+t_n})  \max(G, F_{t_{n+1}}) \mid \mathcal{F}_{t_n} \right]\right\}.
    \label{eq:eventA}
    \end{equation}
    That is, $A_n$ represents surrendering the policy at $t_n$ is optimal, while $A_n^c$ means holding the policy is optimal at $t_n$.
    The optimal stopping time $\tau$ can be obtained from the recursive relation:
    \[
\tau_N=t_N, \tau_n=t_n\, \mathbf{1}_{A_n}+\tau_{n+1}\, \mathbf{1}_{A^c_{t_n}}, \text{ for } 0 \leq n \leq N-1.
    \]
We define the discounted payoff from time $t_n$ onward by the backward recursion
\begin{align}
\Pi_{\tau_N} &:= \max(G, F_{t_N}), \nonumber\\
\Pi_{\tau_n} &:= g_{t_n}\, \mathbf 1_{\{\tau_n=t_n\}}
+ e^{-rh}\Big(\px[h]{x+t_n}\, \Pi_{\tau_{n+1}} + (1-\px[h]{x+t_n})\max(G, F_{t_{n+1}})\Big)\, \mathbf 1_{\{\tau_n>t_n\}},
\label{eq:Pi_recursion}
\end{align}
for $n = N-1, \ldots, 0$.

 The value process (Snell envelope) on the grid is linked with the discounted payoff by
\begin{equation}
U_{t_n} := \sup_{\tau\in\mathcal{T}^N_{[t_n, T]}} \mathbb{E}^{\mathbb{Q}}[\Pi_\tau\mid \mathcal F_{t_n}],
\label{eq:Snell_U}
\end{equation}
where $\mathcal{T}^N_{[t_n, T]}$ denotes the set of stopping times taking values in the discrete set $\{t_n, \cdots, t_N\}$ for $n=0, \cdots, N$.

On the grid $\{t_n\}$, the value process $(U_{t_n})_{n=0, \dots,N}$ of the VA satisfies the relationship
\[
U_{t_n} = \max\{ g_{t_n}, C_{\tau_n}\},
\]
where $C_{\tau_n}$ is defined as the continuation value at time $t_n$ based on the optimal stopping rule $\tau_n$, which is given by
\begin{equation}
C_{\tau_n}:=\mathbb{E}^{\mathbb{Q}}\left[e^{-rh}\Big(\px[h]{x+t_n}\Pi_{\tau_{n+1}}+(1-\px[h]{x+t_n})\max(G, F_{t_{n+1}})\Big)\bigg|\mathcal F_{t_n}\right].
\label{eq: Ctaun}
\end{equation}
Note that in the rest of the paper, for a general stopping rule $\tau$ (not necessarily the optimal one), we also write $\Pi_\tau$ and $C_\tau$ as the discounted payoff and the continuation value obtained by the stopping rule $\tau$.

The LSMC algorithm is a powerful simulation-based method for solving optimal stopping problems. The major challenge is to express the continuation value at time $t_n$, which is the RHS of the inequality in \eqref{eq:eventA}, as a function of the state variables. In Markovian models, \cite{LongstaffSchwartz2001} approximates the continuation value as a polynomial of the state variables. However, in the context of our variable annuity contract, both the equity price model and the mortality model are non-Markovian, and traditional methods fail due to the high dimensionality of the state variable. Therefore, we apply the deep signature method proposed in \cite{bayer2025pricing} to approximate the continuation value, accounting for the non-Markovian features, thereby extending the LSMC algorithm to non-Markovian markets. The details of the deep signature LSMC method are discussed in Sections \ref{sec: State Variable Encoding and Neural Network Structure}-\ref{sec: Fair fee}.

\subsection{State variable encoding and neural network structure}
\label{sec: State Variable Encoding and Neural Network Structure}
We use signature encoding as an input feature for our neural-network-based regression, capturing full-path information. Path signatures summarize sequential data by capturing both the order and interaction of increments along the path. 
Under the non-Markovian framework, there are two benefits to encoding the path with a rough signature. The first is that the rough path is uniquely determined by the signature (see \cite{boedihardjo2016signature}). The second is that the signature characterizes a path/time series of variable length as a fixed-dimensional representation, avoiding the high dimensionality in the time direction. 

To begin with, we start from the classical signature. The classical signature is defined on continuous paths of bounded variation.

\begin{definition}
(\textbf{classical signature and truncated signature}).\\
Let $X: [0,T] \to \mathbb{R}^d$ be a continuous path with bounded variation, denoted by $X:=(X_t^1, \cdots,X_t^d),$ where each $X^i:[0,T]\to \mathbb{R}$ is a real-valued path. For $0\leq s \leq t \leq T$ and an integer $k\geq 1$, define the \textbf{$\text{level-}k$ iterated integrals} of $X$ over $[s,t]$ by
\[
X^{(k)}_{[s, t]}(i_1, \dots, i_k):= \int_{s<u_1<\cdots<u_k<t}dX^{i_1}_{u_1}\cdots dX^{i_k}_{u_k}, i_j\in \{1, \cdots, d\}.
\]
The \textbf{$\text{level-}k$ signature} is a tensor collecting all the $\text{level-}k$ iterated integrals, with each iterated integral representing a combination of the indices $\{i_1, \dots, i_k\}$
\[
X^{(k)}_{[s, t]}:=\left(X^{(k)}_{[s, t]}(1, \dots, 1), X^{(k)}_{[s, t]}(1, \dots, 1, 2), \cdots, X^{(k)}_{[s, t]}(d, \dots, d)\right)\in (\mathbb{R}^d)^{\otimes k}
\]
The \textbf{signature} of $X$ over $[s, t]$ is the infinite collection of signatures of all the levels, denoted by
\[
\text{Sig}^{<\infty}(X)_{[s, t]}:=\left(1, X^{(1)}_{[s, t]}, X^{(2)}_{[s, t]}, \cdots\right) \in \widetilde{T}(\mathbb{R}^d),
\]
where $\widetilde{T}\left((\mathbb{R}^d)\right):=\oplus_{k\geq0}^\infty (\mathbb{R}^d)^{\otimes k}$ is the tensor algebra.\\
For an integer $K \geq 1$, the \textbf{truncated signature up to level $K$} is 
\[
\text{Sig}^{\leq K}(X)_{[s,t]}:=\left(1,X^{(1)}_{[s,t]}, \dots,X^{(K)}_{[s,t]}\right) \in \widetilde{T}^{\leq K}(\mathbb{R}^d),
\]
where $\widetilde{T}^{\leq K}(\mathbb{R}^d):=\oplus_{k=0}^K(\mathbb{R}^d)^{\otimes k}.$
\label{def:classical_sig}
\end{definition}

We denote the signature dimension by $d_K:=d^0+d^1+\cdots+d^K$. The classical signature in Definition \ref{def:classical_sig} is defined via Riemann-Stieltjes integration, and is well-defined for paths with bounded variation. However, in stochastic calculus, most processes don't have bounded variation, such as Brownian motion and fractional Brownian motion. To extend the signature to the non-bounded variation process, we first introduce the following two definitions. The first is the $\alpha$-Hölder continuity. For $\alpha \in (0,1)$, a path $X$ is called $\alpha$-Hölder-continuous if
\begin{equation*}
    ||X||_{\alpha-\text{Höl}}:=\sup \limits_{0\leq s<t\leq T} \frac{||X_t-X_s||_{\mathbb{R}^d}}{|t-s|^{\alpha}}<\infty,
\end{equation*}
where $||\cdot||_{\mathbb{R}^d}$ is the Euclidean distance on $\mathbb{R}^d$. We use $C^{\alpha-\text{Höl}}([0,T];\mathbb{R}^d)$ to denote the space of $\alpha$-Hölder-continuous paths. The second is the finite $p$-variation process. A path is called a finite $p$-variation process if for some $p>0$
\begin{equation*}
    ||X||_{\text{p-var}}:=\left(\lim \limits_{||\Pi||\rightarrow 0} \sum \limits_{}||X_{t_{i+1}}-X_{t_i}||_{\mathbb{R}^d}^p\right)^{\frac{1}{p}}<\infty,
\end{equation*}
where $\Pi = \{0=t_0<t_1<...<t_n=T\}$ is a partition of $[0,T]$ and $||\Pi||=\max \limits_{0\leq j \leq n-1}(t_{j+1}-t_j)$. Moreover, we let $C^{\text{p-var}}([0,T];\mathbb{R}^d)$ denote the space for the p-variation process. The relationship between these two notions is straightforward. If a process is $\alpha$-Hölder-continuous, then it has finite $p$-variation for any $p\ge1/\alpha$. The converse is not true, as the $p$-variation process need not be continuous (e.g., step functions); see Section 5.1.1 in \cite{friz2010multidimensional} for details. 

In rough theory, $L=\lfloor 1/\alpha\rfloor$, where $\lfloor \cdot \rfloor$ is the floor function, is a key parameter that determines the level of signature required to represent the path. Specifically, following the Lyons' extension theorem (see Theorem 2.2.1 in \cite{lyons1998differential}), one can write the full level signature by just defining the first $L=\lfloor 1/\alpha\rfloor$ level signature. Following  \cite{BayerPelizzariSchoenmakers2025},
for $L\in \mathbb{N}$ and any two-parameter function on $\widetilde{T}^{\leq L}(\mathbb{R}^d)$
\[
  (s,t)\longmapsto \mathbf{X}_{[s,t]}
   = \bigl(1, \mathbf{X}^{(1)}_{[s,t]}, \dots, \mathbf{X}^{(L)}_{[s,t]}\bigr)
   \in \widetilde{T}^{\le L}(\mathbb{R}^d),
\]
define the norm
\[
||| \mathbf{X} |||_{(\alpha,L)}:=\max_{1\leq l \leq L }\left(\sup_{0 \leq s \leq t \leq T}\dfrac{||\mathbf{X}_{s,t}^{(l)}||_{\mathbb{R}^d}}{|t-s|^{l\alpha}}\right),
\]
where $||\cdot||_{\mathbb{R}^d}$ is the Euclidean distance on $\mathbb{R}^d$. We consider an $\alpha$–Hölder continuous path $X:[0,T]\to\mathbb{R}^d$ with
$\alpha\in(0,1)$ and set $L=\lfloor 1/\alpha\rfloor$. We denote by
$\mathcal{C}^\alpha([0,T];\mathbb{R}^d)$ the space of geometric
$\alpha$–Hölder rough paths $\mathbf{X}$ over $\mathbb{R}^d$, which is the  $|||\, \cdot\,|||_{(\alpha,L)}$ closure of $L$-level truncated signature of Lipschitz continuous path $X$ where $L=\lfloor 1/\alpha\rfloor$. Subsequently, there exists a sequence of Lipschitz paths
        $X^n:[0,T]\to\mathbb{R}^d$ whose truncated signatures
        $\mathbf{X}^n := \text{Sig}^{\le L}(X^n)$ converge to $\mathbf{X}$ in the
        rough-path norm, i.e.
        \begin{equation}
          |||\mathbf{X}^n - \mathbf{X}|||_{(\alpha, L)} \;\to\; 0
          \quad\text{as }n\to\infty, \label{signature_smooth_approx}
        \end{equation}
where truncated signature $\mathbf{X}^n$ is defined in a Riemann–Stieltjes sense.

We cannot directly define the signature of the process $X_t$ with $\alpha$-Hölder continuity when $\alpha\neq 0.5$, since it is not a semimartingale. To circumvent this difficulty, we should lift the path into a geometric rough path, whose signature can be constructed as the limit of the smooth approximation \eqref{signature_smooth_approx}. In the numerical simulation, we compute the signature of a rough path by treating the discrete data as a piecewise-linear path, with $n$ denoting the number of time intervals. When $n$ goes to infinity, the computed signature converges to the true signature for the rough path. From the arguments above, we arrive at the definition for the rough-path signature.

\begin{definition}
\label{def:rough_path_signature}
    (\textbf{rough-path signature})
    Let $X:[0,T]\rightarrow\mathbb{R}^d$ be a continuous path. Assume that the time-augmented path $(t,X_t)$ admits geometric $\alpha-$Hölder rough-path lifts $\widehat{\mathbf{X}}$. Then, by Lyons' extension theorem (see Theorem 2.2.1 in \cite{lyons1998differential}), there exists a unique extension
    \[
    \mathbf{X}^{<\infty}_{[s,t]}:=\left(1, \mathbf{X}_{[s,t]}^{(1)}, \cdots, \mathbf{X}_{[s,t]}^{(L)}, \mathbf{X}_{[s,t]}^{(L+1)}, \cdots\right), \text{ } 0\leq s \leq t \leq T,
    \]
where $L=\lfloor 1/\alpha\rfloor$, satisfying Chen's relation at all levels and $\sup_{s\leq s^\prime \leq t^\prime \leq t} \dfrac{||\mathbf{X}^{(k)}_{t^\prime}-\mathbf{X}^{(k)}_{s^\prime}||_{\mathbb{R}^d}}{|t^\prime-s^\prime|^{k\alpha}}<\infty, \forall k\geq0.$
we call $\mathbf{X}^{<\infty}_{[s,t]}$ the (rough-path) signature of $X$ and write
\[
\text{Sig}^{<\infty}(X)_{[s,t]}:=\mathbf{X}^{<\infty}_{[s,t]}.
\]
The truncated rough-path signature up to level $K$ is 
\[
\text{Sig}^{\leq K}(X)_{[s,t]}:= \left(1, \mathbf{X}^{(1)}_{[s,t]}, \cdots, \mathbf{X}^{(K)}_{[s,t]}\right)\in \widetilde{T}^{\leq K}(\mathbb{R}^d).
\]
\end{definition}

Lyons' extension theorem ensures that the iterated integration of rough paths makes sense, and the rough path signature shares the same property as the classic signature of smooth paths. Hence, we use the same notation $\text{Sig}^{<\infty}(\cdot)$ and $\text{Sig}^{\leq K}(\cdot)$ to denote the full and truncated signatures of both smooth and rough paths.

Back to our model, the stock volatility $V_t$ follows a rough Heston model with Hurst parameter $H_S \in (0,1/2)$. In contrast, the force of mortality $\mu_{x+t}$ follows a smoother path with a Hurst parameter $H_m \in (1/2,1)$. The roughness of the time-augmented path $\widehat{X}_t = (t, V_t, \mu_{x+t})$ is determined by its roughest part (one can obtain this result from Definition 5.50 in \cite{friz2010multidimensional}). More accurately, $t$ is 1-variation; $V_t$ is $p_1$-variation with $p_1>1/H_S$ and $H_S<1/2$, so $p_1>2$ and $V_t$ is rougher than the Brownian motion; $\mu_{x+t}$ is $p_2$-variation with $p_2>1/H_m$ and $H_m=0.703923>1/2$, so $1<p_2<2$ and $\mu_{x+t}$ is smoother than a Brownian motion. In general, the time-augmented path $\widehat{X}_t = (t, V_t, \mu_{x+t})$ is $p$-variation with $p=\max(1,p_1,p_2)=p_1$, which is the same as the stock volatility path. In practice, the Hurst parameter of stock volatility is typically estimated at around $H_S = 0.1$. It means one needs at least level $L=\lfloor 1/H_S \rfloor=10$ to define the whole rough-path signature in Definition \ref{def:rough_path_signature}. To save computation effort, the existing literature usually chooses a small truncation level, such as $K=2 ~\text{or}~ 3$. The reason is that these low-level signatures have already captured most of the path information, enabling them to fit the implied volatility smile with a small absolute error (see \cite{cuchiero2023signature}, \cite{BayerPelizzariSchoenmakers2025}, and \cite{jaber2025signature}). In this paper, we determine the signature truncation level $K$ from the hyperparameter tuning.

\begin{figure}
\centering

\tikzset{%
  every neuron/.style={
    circle,
    draw,
    minimum size=17pt
  },
  neuron missing/.style={
    draw=none,
    scale=2,
    text height=0.333cm,
    execute at begin node=\color{black}$\vdots$
  }
}

\begin{tikzpicture}[x=1.5cm, y=1.5cm, >=stealth]

  \node [every neuron/.try] (input-1) at (0,2.5) {};
\node [every neuron/.try] (input-2) at (0,0.5) {};


  \foreach \m [count=\y] in {1,missing,missing,2}
    \node [every neuron/.try, neuron \m/.try] (hidden1-\m) at (1.5,5.0-\y*1.5) {};

  \foreach \m [count=\y] in {1,missing,missing,2}
    \node [every neuron/.try, neuron \m/.try] (hidden2-\m) at (3,5.0-\y*1.5) {};

  \foreach \m [count=\y] in {1,missing,missing,2}
    \node [every neuron/.try, neuron \m/.try] (hidden3-\m) at (4.5,5.0-\y*1.5) {};

  \foreach \m [count=\y] in {1,missing,missing,2}
    \node [every neuron/.try, neuron \m/.try] (hidden4-\m) at (6,5.0-\y*1.5) {};

  \node [every neuron/.try] (output-1) at (7.5,2.2-1) {};


  \node [above] at (input-1.north) {$\log S_t$};
  \node [above] at (input-2.north) {$\text{Sig}^{\leq3}\left((s, V_s, \mu_{x+s})\right)_{[0,t]}$};

  \node [above] at (hidden1-1.north) {$H^{(1)}_{1}$};
  \node [above] at (hidden1-2.north) {$H^{(1)}_{\widehat{p}}$};

  \node [above] at (hidden2-1.north) {$H^{(2)}_{1}$};
  \node [above] at (hidden2-2.north) {$H^{(2)}_{\widehat{p}}$};

  \node [above] at (hidden3-1.north) {$H^{(3)}_{1}$};
  \node [above] at (hidden3-2.north) {$H^{(3)}_{\widehat{p}}$};

  \node [above] at (hidden4-1.north) {$H^{(4)}_{1}$};
  \node [above] at (hidden4-2.north) {$H^{(4)}_{\widehat{p}}$};

  \node [above] at (output-1.north) {Continuation Value};


  \foreach \i in {1,2}
    \foreach \j in {1,2}
      \draw[->] (input-\i) -- (hidden1-\j);

  \foreach \i in {1,2}
    \foreach \j in {1,2}
      \draw[->] (hidden1-\i) -- (hidden2-\j);

  \foreach \i in {1,2}
    \foreach \j in {1,2}
      \draw[->] (hidden2-\i) -- (hidden3-\j);
    
  \foreach \i in {1,2}
    \foreach \j in {1,2}
      \draw[->] (hidden3-\i) -- (hidden4-\j);
      
  \foreach \i in {1,2}
    \draw[->] (hidden4-\i) -- (output-1);

  \foreach \l [count=\x from 0] in {Input, Hidden 1, Hidden 2, Hidden 3, Hidden 4, Output}
    \node [align=center, above] at (\x*1.5,4.5) {\l};

\end{tikzpicture}

\caption{Neural network with structure ``$41-64-64-64-64-1$''.}
\label{fig:NN_structure}
\end{figure}
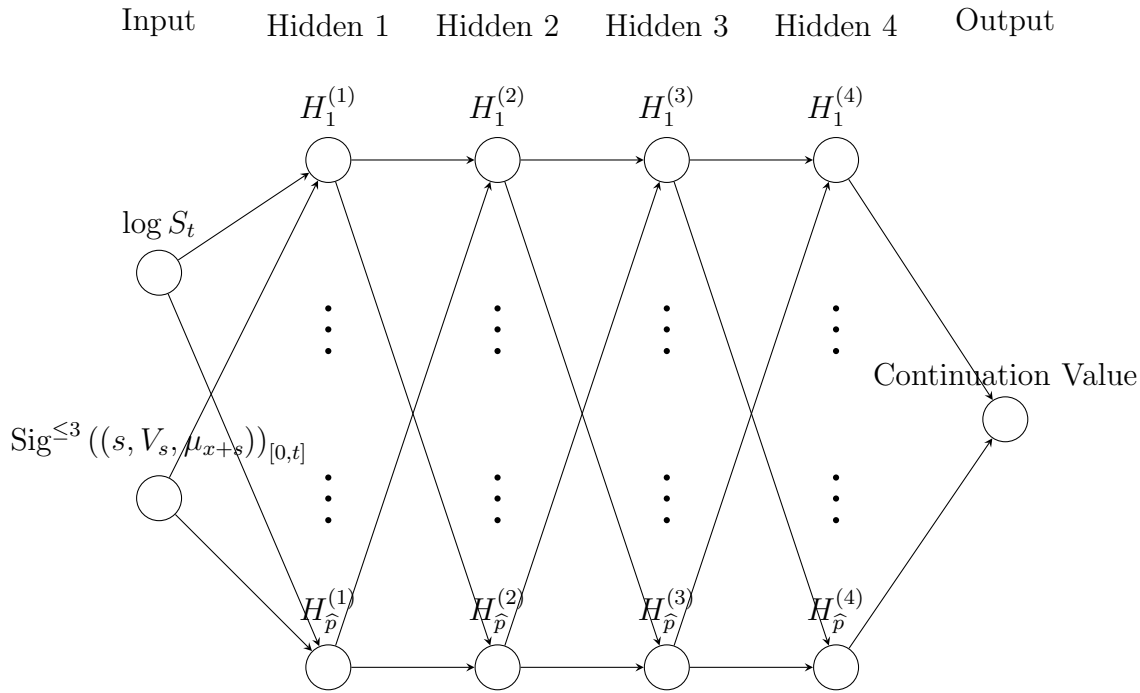

Next, we can apply a deep signature LSMC algorithm to estimate the VA continuation value, and the neural network structure is designed as (see Figure \ref{fig:NN_structure}):
\begin{itemize}
    \item The input to the regression neural network is the path signature of the time‑augmented path $(t, V_t, \mu_{x+t})$, concatenated with a local feature $\log{S_t}$.
    \item The signature is computed numerically using the \texttt{iisignature} library. We take the truncated signature up to level $3$, and since the time-augmented path $(t, V_t, \mu_{x+t})$ is 3-dimensional, the truncated signature is of dimension $40$ ($3^0+3^1+3^2+3^3=40$).
    \item The neural network consists of 4 fully connected layers, each with 64 nodes. Following \cite{LapeyreLelong2021} and \cite{bayer2025pricing}, \texttt{LeakyReLU} activations with a negative slope of 0.3 are applied, as \texttt{LeakyReLU} activation function prevents the ``dying neuron problem" and improves gradient flow.
    \item The final output layer is a single neuron that returns the predicted continuation value.
\end{itemize}

All parameters above, the signature truncation level, the neural network width and depth, training batch size, and activation function, are determined by a hyperparameter tuning process. We tested numerous hyperparameter combinations and selected the one that produced the highest VA price on the test set. Table \ref{table:hyper} lists the candidate and the selected parameters. In contrast to traditional deep learning tasks, our simulation-based regression problem benefits from a larger batch size as it reduces the variance of stochastic gradients and yields a more stable approximation of the conditional expectation. This is particularly important in the backward induction procedure, since regression errors at one exercise date accumulate to earlier dates through the optimal stopping recursion. Moreover, unlike standard supervised learning problems, our training data are generated by Monte Carlo simulation rather than drawn from a fixed finite dataset. Hence, our sample size is not limited by a fixed empirical dataset, and it is natural to use a relatively large batch size to improve the stability of the regression.

\begin{table}[h!]
\centering
\caption{Hyperparameter tuning}
\label{table:hyper}
\begin{tabular}{lll}
\hline
\textbf{Hyperparameter} & \textbf{Candidates} & \textbf{Selected} \\
\hline
Signature level $K$ & $2, 3, 4, 5$ & $3$ \\
Hidden layers & $2, 3, 4$ & $4$ \\
Width & $32, 64, 128$ & $64$ \\
Batch size & $2^8, 2^{10}, 2^{12}$ & $2^{12}$\\
Activation & ReLU, LeakyReLU & LeakyReLU \\
\hline
\end{tabular}
\end{table}

\subsection{Deep signature LSMC method}
\label{sec: Deep signature LSMC method}
\paragraph{State Variables.}  
At each decision time $t_n$, the state variable includes all information needed to predict the future evolution of both financial and mortality risks
\begin{equation*}
    \Big( t_n, S_{t_n}, V_{[0, t_n]}, \mu_{[x, x+t_n]}  \Big),
\end{equation*}
where $t_n$ is current time, $S_{t_n}$ is the current equity value, $V_{[0, t_n]}$ is the path of volatility up to $t_n$, and $\mu_{[x, x+t_n]}$ is the path of the individuals' force of mortality from age $x$ to age $x+t_n$.
Due to the special form of \eqref{eq: heston}, $S_t$ has a partial Markovian nature in Brownian motion $W_t$. In other words, the evolution of $S_t$ only depends on current values of price and variance $S_t, V_t$. Therefore, we only need to include the current value $S_{t_n}$ into the state variables. In contrast, the rough Heston variance $V_t$ and the force of mortality $\mu_{x+t}$ are non-Markovian: their future dynamics depend on the entire past trajectory through the Volterra integration term. As a result, we include the entire historical path of $V_t$ in the signature calculation, while only the current stock price is used in the neural network input.

\paragraph{Deep signature LSMC algorithm.}  
The deep signature LSMC algorithm is implemented by backward induction over the possible early surrender dates $\{t_n\}_{n=0}^N$. At each time $t_n$, the policyholder chooses whether to surrender the contract or to continue. The continuation value at time $t$ is defined as the conditional expectation of the survival and death benefits, weighted by the corresponding survival and death probabilities. In particular, the survival benefit is the discounted fund value under the optimal surrender strategy. The death benefit is calculated as the discounted payoff at the policyholder's death. 

The key steps of deep signature LSMC algorithm are as follows:

\begin{enumerate}
    \item \textbf{Simulation of sample paths:} Simulate $M$ independent sample paths of the joint process
    $$
    \left\{ (S_{t_k}^{(i)}, V_{t_k}^{(i)}, \mu_{x
    +t_k}^{(i)}) \right\}_{k=0}^N, \quad i = 1, \dots, M,
    $$
    under the measure $\mathbb{Q}$. Each path includes the full history of volatility and mortality up to each decision time.
    
    \item \textbf{Initialization at maturity:}  
    At the final time $t_N = T$, we initialize the optimal surrender time of each path $i$ at the terminal time
    \[
    \tau^{(i)}_{t_N}=t_N,
    \]
    and set the optimal payoff for each path $i$ at current time $t_n$, denoted by $\Pi_{\tau_n^{(i)}}^{(i)}$
    \[
\Pi_{\tau_N^{(i)}}^{(i)} = \max \left( G, F_{t_N}^{(i)} \right).    
    \]

    \item \textbf{Backward induction:}  
    For $n = N-1, N-2, \ldots, 1$, conduct the following steps:
    \begin{enumerate}
        \item \textbf{Surrender payoff}: For each simulated path $i$, compute the immediate surrender payoff at current time $t_n$:
        \[
        g_{t_n}^{(i)} = 
            (1 - \kappa_{t_n}) F_{t_n}^{(i)}.
        \]
        \item \textbf{Compute realization of continuation value}: We denote the realization of the continuation value of path $i$ by $C_{\tau_n^{(i)}}^{(i)}$. The realized continuation value is the discounted expected payoff of the survival benefit and the death benefit:
        \begin{equation}
        {C}_{\tau_n^{(i)}}^{(i)} = e^{-r h} \left( \px[h]{x+t_n}^{(i)} \, \Pi_{\tau_{n+1}^{(i)}}^{(i)}
        +   \, (1-\px[h]{x+t_n}^{(i)})\max \left(G, F_{t_{n+1}}^{(i)}\right) \right),
        \label{eq: DPP_discrete}
        \end{equation}
        where $\px[h]{x+t_n}^{(i)}$ is the realized survival probability for the $i-$th path based on the force of mortality path $\mu_{x+t}^{(i)}$.

        \item \textbf{Regression step:}
    Let $\mathbf{x}_{t_n}\in \mathbb{R}^{d_K+1}$ denote the fixed-dimensional feature vector constructed from the information available up to time $t_n$ using signature encoding and local features (see Section~\ref{sec: State Variable Encoding and Neural Network Structure}). We approximate the continuation value $C_{\tau_n}$ by a neural network $\Phi(\phi_n, \mathbf{x}_{t_n})$, where $\phi_n$ denotes the network parameters at time $t_n$
    \[
    C_{
    \tau_n} \approx \Phi(\phi_n, \mathbf{x}_{t_n}).
    \]  

        Subsequently, the optimal neural network parameters $\phi_n^*$ are obtained by minimizing the Mean Square Error (MSE) between the continuation value estimated by the neural network and the realized policy value of each path, $C_{\tau_n^{(i)}}^{(i)},$ for $i=1, \cdots, M,$
        \[
        \phi_n^* = \arg\min_{\phi_n} \dfrac{1}{M}\sum_{i=1}^{M} \left(C_{\tau_n^{(i)}}^{(i)}-\Phi(\phi_n, \mathbf{x}^{(i)}_{t_n})\right)^2 .
        \]\\
        Note that our design is slightly different from \cite{LongstaffSchwartz2001}, as they only include ``in-the-money" paths in the regression step, but our algorithm includes all the paths for regression. This is because in the context of variable annuity, the immediate surrender payoff is always positive.
        \item \textbf{Optimal decision:}  
        Compare the surrender payoff and predicted continuation value. We denote by $A_n$ the event that the surrender payoff is higher than the predicted continuation value
        \[
        A^{(i)}_{t_n}:=\left\{g_{t_n}^{(i)} \ge \Phi(\phi_n^*, \mathbf{x}^{(i)}_{t_n})\right\}.
        \]
        For each path $i$, update the realized policy value by taking the maximum of the immediate surrender value and the estimated continuation value, and update the optimal surrender time $\tau^{(i)}_{t_n}$ if early surrender is optimal

        \begin{equation}
\Pi^{(i)}_{\tau_n^{(i)}}
=
g_{t_n}^{(i)}\, \mathbf{1}_{ A^{(i)}_{t_n}}
+
C^{(i)}_{\tau_n^{(i)}}\, \mathbf{1}_{( A^{(i)}_{t_n})^c},
\qquad
\tau^{(i)}_{t_n}
=
t_n\, \mathbf{1}_{A^{(i)}_{t_n}}
+
\tau^{(i)}_{t_{n+1}}\, \mathbf{1}_{(A^{(i)}_{t_n})^c}.
\label{eq: update}
\end{equation}

        
    \end{enumerate}
    \item \textbf{Price of VA policy:} Backward iterate until $n=0$ and calculate $\Pi^{(i)}_{\tau_0^{(i)}}$ by (\ref{eq: update}).
        The price of the VA policy is calculated as the average of the Monte Carlo samples
        \[
        \dfrac{1}{M}\sum_{i=1}^M\Pi_{\tau_0^{(i)}}^{(i)}.
        \]
\end{enumerate}

The optimal surrender time for each path, $\tau_0^{(i)}$, can also be obtained as the first time $t_n$ at which the immediate payoff equals or exceeds the predicted continuation value. i.e., 
\[
\tau_0^{(i)}=\inf\left\{t_n \in [0,T] :g_{t_n}\ge \Phi(\phi_n^*, \mathbf{x}^{(i)}_{t_n})\right\}.
\]

The LSMC algorithm is prone to the so-called ``look-ahead bias" (see \cite{WooLiuChoi2024LOOLSM}). This occurs because the same simulated paths are used both to learn the surrender strategy and to value the policy under that strategy. In each regression step, the continuation value on a given path is partly fitted to that path’s own future payoff. As a result, the algorithm is more likely to “decide” to continue exactly on those paths where the future payoff (seen in the simulation) turns out to be high, and to exercise when the future payoff would be low, making the algorithm systematically overestimate the policy value.

A common practice to remove the ``look-ahead" bias is using a “two‑pass” approach: use one set of sample paths to learn the surrender strategy (use the training path set to study the optimal weights and biases $\phi^*$ of the neural network) and apply the strategy to another independent set of sample paths (use the testing path set to compute the VA price).  Therefore, to avoid ``look-ahead bias", we generate a testing path set to compute the true VA price
\[\left(t, \widetilde{S}_t^{(i)}, \widetilde{V}_t^{(i)}, \widetilde{\mu}_{x+t}^{(i)}\right), t=t_0, \cdots,t_N, i=1, \cdots, \widetilde{M},
\]
where the superscript $\widetilde{\cdot}$ means the test samples. We set the sample path number for both training and test sets, i.e. $\widetilde{M}=M$. Next, the estimated VA price at time $0$ is the sample average of the expected discounted payoff of the Monte Carlo paths
\[
\text{Price}= \dfrac{1}{\widetilde{M}}\sum_{i=1}^{\widetilde{M}} \left\{ {}_{\widetilde{\tau}_0^{(i)}}\widetilde{p}_{x}^{(i)} \, e^{-r\widetilde{\tau}_0^{(i)}}\, \widetilde{g}_{\widetilde{\tau}_0^{(i)}}^{(i)}+\sum_{t_j=t_0,t_1, \cdots, \widetilde{\tau}_0^{(i)}-h} {}_{t_j}\widetilde{p}_{x}^{(i)} \, (1-{}_h\widetilde{p}_{x+t_j}^{(i)}) \, e^{-rt_j} \, \max(G, \widetilde{F}_{t_j}^{(i)}) \right\},
\]
where $\widetilde{\tau}_0^{(i)}=\inf\left\{t_n \in [0,T]:\widetilde{g}_{t_n}^{(i)}\ge\Phi(\phi_n^*, \widetilde{\mathbf{x}}^{(i)}_{t_n})\right\}$, $\phi_n^*$ are the optimal weights and biases obtained from the training path set $\mathbf{x}^{i}$, $\widetilde{F}_t=e^{-ct}\widetilde{S}_t$ is the account value, $\widetilde{g}_{t_n}^{(i)}$ is the payoff of the $i$--th testing sample path at time $t_n$,  $\widetilde{\mathbf{x}}^{(i)}_{t_n}$ is the state variables of the $i-$th testing sample path at time $t_n$, and the realized survival probability ${}_\cdot\widetilde{p}_{\cdot}^{(i)}$ is calculated based on the force of mortality paths $\widetilde{\mu}_{x+t}^{(i)}$:
\[
{}_{t_n}\widetilde{p}_{x}^{(i)}\approx\exp \left(-\frac{h}{2}\sum_{k=0}^{n-1}\widetilde{\mu}_{z+t_k}^{(i)}+\widetilde{\mu}_{z+t_{k+1}}^{(i)}\right).
\]

The deep signature LSMC algorithm provides a practical and robust method for valuing VA contracts and computing fair fees under more realistic, path-dependent models, where classical methods such as the finite-difference method are infeasible.

\subsection{Fair fee}
\label{sec: Fair fee}
In variable annuity contracts, the insurer provides the policyholder with guaranteed benefits and the right to early surrender. However, these guaranteed benefits do not come for free; they are financed by fees continuously deducted from the VA account. For a given fee level $c\ge 0$, let $U_0(c)$ denote the time-zero value of the VA under the optimal surrender strategy (defined via the optimal stopping problem \eqref{eq: Ut}). The fair fee (denoted by $c^*$) is therefore the fee level that exactly compensates the insurer for providing the embedded guarantees, and it makes the VA contract actuarially fair at inception, namely
\[
U_0(c^*)=F_0,
\]
where $F_0$ is the initial premium (and initial account value). As indicated in \cite{Shen2016IME}, this creates a nested problem: Changing the fee rate $c$ alters the account value dynamics and surrender incentives (charging a higher fee gives the policyholder greater incentives to surrender the VA early to avoid fee costs), which in turn affects the contract value and hence the fair fee itself.

In the traditional Markovian setting, the fair fee can be solved using variational inequality methods. However, these approaches become impractical in our non-Markovian setting due to the path dependence in both market and mortality models. We therefore compute $c^*$ numerically using bisection. For any fixed fee level $c \geq 0$, we first evaluate the corresponding VA price $U_0(c)$ by solving the optimal stopping problem (see Section~3.2 for details). Since higher fees reduce the contract value, $U_0(c)$ is decreasing in $c$, making the bisection method well-suited. Given a fixed initial premium $F_0$, we start from an interval $[c_L, c_U]$ such that $U_0(c_L)\ge F_0\ge U_0(c_U)$. We iteratively set $c_{\mathrm{mid}}=(c_L+c_U)/2$ and update the upper and lower bounds depending on whether $U_0(c_{\mathrm{mid}})$ is larger or smaller than $F_0$ until $|c_U-c_L|$ falls below a prescribed tolerance.

\section{Convergence analysis of the LSMC algorithm for the optimal surrender strategy}
\label{sec: convergence}
In this section, we analyze the convergence of the deep signature LSMC algorithm for valuing the variable annuity with an early surrender option. We consider a finite time grid
\[
0=t_0<t_1<\cdots<t_N=T,
\]
and first analyze the convergence for the discretized optimal stopping problem. The discretization error is discussed at the end.

We assume the joint process $(S_t, V_t, \mu_{x+t})$ satisfies the equity price and mortality models in Section \ref{sec: problem_statement}. For an $\alpha-$Hölder continuous path $X\in\mathbb{R}^d$, we denote by $\widehat{\mathcal{C}}^\alpha([0, T];\mathbb{R}^{d+1})$ the space of geometric $\alpha-$Hölder rough-path lifts of $(\widehat{X_t})=(t, X_t)$, and make the following assumption. In the following parts, we extend the convergence proof in \cite{BayerPelizzariSchoenmakers2025} from the linear signature method to the deep signature method. The following assumption is needed to guarantee the regularity of the time-augmented path $\widehat{X}_t=(t, V_t, \mu_{x+t})$. 

\begin{assumption}
(i) $\alpha \in (0,1)$ such that $1/\alpha \notin \mathbb{N}$;\\
(ii) The Borel space $(\widehat{\mathcal{C}}^\alpha,\mathcal{B}(\widehat{\mathcal{C}}^\alpha))$  has a measure $\widehat{\mu}$ such that
\begin{equation*}
    \widehat{\mu}(\mathcal{C}^\alpha)<\infty ~~\text{and}~~ \widehat{\mu}(\mathcal{C}^{\alpha}/\mathcal{C}^{\beta})=0, ~\text{for}~ \forall \alpha<\beta<\frac{1}{\lfloor 1/\alpha \rfloor}.  
\end{equation*}
\label{assump: regularity}
\end{assumption}
\vspace{-0.5cm}
\begin{lemma}
\label{lemma:universal_approx}
Under Assumption \ref{assump: regularity}, for each $n\in \{0, \cdots, N-1\}$, there exists $\phi_n^{\widehat{p}, K} \in \Theta_{\widehat{p}}$, the neural network parameters with neural network width $\widehat{p}$ and model input of path signature truncated at level $K$, such that
\[
\mathbb{E}\left[|C_{\tau_n}-\Phi_{\widehat{p}}(\phi_n^{\widehat{p}, K}, \text{Sig}^{\leq K}(\widehat{X})_{[0, t_n]}))|^2\right]\xrightarrow[\widehat{p}, K\to\infty]{} 0,
\]
where $C_{\tau_n}$ is defined in (\ref{eq: Ctaun}).
\noindent
\begin{proof}
By Lemma 2.4 of \cite{BayerPelizzariSchoenmakers2025}, there exists a measurable function $f$ of the stopped path $(\widehat{X}_t)_{0\leq t\leq t_n}$ such that 
$C_{\tau_n}=f((\widehat{X}_t)_{0\leq t\leq t_n})$ almost everywhere. Hence, it belongs to the $L^2-$space of path functionals considered in their Theorem 2.8. Next, using their Theorem 2.8, we first approximate the function $f((\widehat{X}_t)_{0\leq t\leq t_n})$ by linear functionals of the signature, i.e., for any $\varepsilon>0$, we can find a linear function of the signature $\mathcal{l}$ such that
\[
|| \mathcal{l}(\text{Sig}^{<\infty}(\widehat{X})_{[0, t_n]})-C_{\tau_n}||_{L^2}<\varepsilon/3.
\]
and we can choose a sufficiently large truncation level $K$ such that
\[
|| \mathcal{l}(\text{Sig}^{<\infty}(\widehat{X})_{[0, t_n]})-\mathcal{l}(\text{Sig}^{\leq K}(\widehat{X})_{[0, t_n]}) ||_{L^2}<\varepsilon/3.
\]
Finally, we approximate those linear functionals by using neural networks of truncated signatures. By Assumption \ref{ass:A1} and the universal approximation of neural network(see \cite{LeshnoLinPinkusSchocken1993} and \cite{bayer2025pricing}), for the fixed truncation level $K$ and any $\varepsilon>0$, there exists $\widehat{p}$ and $\phi^{\widehat{p}, K}\in \Theta_{\widehat{p}}$ such that 
\[
|| \Phi_{\widehat{p}}(\phi^{\widehat{p}, K}, \text{Sig}^{\leq K}(\widehat{X})_{[0, t_n]}) - \mathcal{l}(\text{Sig}^{\leq K}(\widehat{X})_{[0, t_n]}) ||_{L^2}<\varepsilon/3.
\]
Combining the results, we show that for any $\varepsilon>0$ there exists $\widehat{p}, K$ large enough such that 
\[
|| \Phi_{\widehat{p}}(\phi^{\widehat{p}, K}, \text{Sig}^{\leq K}(\widehat{X})_{[0, t_n]})- C_{\tau_n}||_{L^2}<\varepsilon.
\]
\end{proof}
\label{lemma: approx}
\end{lemma}

Recall the discretized formula for continuation value \eqref{eq: DPP_discrete}, we can define the optimal stopping times recursively
\begin{align*}
    \tau_N&=t_N, \\
    \tau_n&=t_n \mathbf{1}_{\{g_{t_n}\ge C_{\tau_n}\}}+\tau_{n+1}\mathbf{1}_{\{g_{t_n}< C_{\tau_n}\}}, \text{ }n=0, \cdots,N-1.
\end{align*}
For fixed neural network width $\widehat{p}$ and signature truncation level $K$, we approximate the optimal stopping times $\tau_n$ by $\tau_n^{\widehat{p}, K}$. Moreover, $C_{\tau_n^{\widehat{p}, K}}$ is the continuation value at time $t_n$ based on $\tau_n^{\widehat{p}, K}$. We use $\mathbf{x}_{t_n}^K$ to denote the feature vector (truncated signature up to level $K$ and local features) at $t_n$. Define $\phi^{\widehat{p}, K}$ as the optimal neural network parameter with width $\widehat{p}$ and truncated signature level $K$ at time $t_n$, which is the solution to the optimization problem
\begin{equation}
    \phi^{\widehat{p}, K}_n:=\arg \inf_{\phi\in\Theta_{\widehat{p}}} ||\Phi_{\widehat{p}}(\phi, \mathbf{x}_{t_n}^K)-C_{\tau_n^{\widehat{p}, K}}||_{L^2},
    \label{eq: nn_param_alpha}
\end{equation}
Subsequently, we obtain the approximating sequence of stopping times
\begin{align*}
    \tau_N^{\widehat{p}, K}&=t_N, \\
    \tau_n^{\widehat{p}, K}&=t_n \mathbf{1}_{\{g_{t_n}\ge \Phi_{\widehat{p}}(\phi^{\widehat{p}, K}_n, \mathbf{x}_{t_n}^K)\}}+\tau_{n+1}^{\widehat{p}, K}\mathbf{1}_{\{g_{t_n}< \Phi_{\widehat{p}}(\phi^{\widehat{p}, K}_n, \mathbf{x}_{t_n}^K)\}}, \text{ }n=0, \cdots,N-1.
\end{align*}
The following proposition shows that the policy payoff converges as the neural network class becomes richer (by increasing the width $\widehat{p}$) and as the signature truncation level increases.

\begin{proposition}
For all $n=0, \cdots,N$, we have 
\[
\lim_{\widehat{p}, K\to \infty} \mathbb{E}[\Pi_{\tau^{\widehat{p}, K}_n}\mid\mathcal{F}_{t_n}]=\mathbb{E}[\Pi_{\tau_n}\mid \mathcal{F}_{t_n}]\text{ in } L^2.
\]

\begin{proof}
    Inspired by Proposition~3.3 in \cite{BayerPelizzariSchoenmakers2025}, we prove the convergence by extending their linear projection of path signature to the deep signature regressors, together with the deep signature approximation result from Lemma \ref{lemma:universal_approx}. \\
\textbf{Base Case.} For $n=N,$ we set $\tau_N^{\widehat{p}, K}=\tau_N=t_N.$ The claim is trivial.\\
\textbf{Induction Step.} \\
\emph{Step 1 (Definition of exercise events and stopping times).}\\
Assume the statement holds for $n+1$. Define the ``true" and approximated surrender decision as 
\[
A_n:=\{g_{t_n}\geq C_{\tau_n}\}, ~A_n(\widehat{p}, K):=\{g_{t_n} \geq \Phi_{\widehat{p}}(\phi_n^{\widehat{p}, K}, \mathbf{x}_{t_n}^K)\}.
\]
The surrender strategy is expressed by the recursive relation
\[
\tau_n=t_n\mathbf{1}_{A_n}+\tau_{n+1}\mathbf{1}_{A_n^c}, ~\tau_n^{\widehat{p}, K}=t_n\mathbf{1}_{A_n(\widehat{p}, K)}+\tau_{n+1}^{\widehat{p}, K}\mathbf{1}_{A_n(\widehat{p}, K)^c}.
\]
\emph{Step 2 (Calculate the difference between the ``true" value and deep signature based value).}\\
By conditioning on survival/death, we write the continuation value at time $t_n$ based on the recursive relationship: 
\[
C_{\tau_n^{\widehat{p}, K}}=e^{-rh}\left\{\mathbb{E}[\px[h]{x+t_n}\Pi_{\tau_{n+1}^{\widehat{p}, K}} \mid \mathcal{F}_{t_n}]+\mathbb{E}[(1-\px[h]{x+t_n})\max(F_{t_{n+1}},G) \mid \mathcal{F}_{t_n}]\right\},
\]
\[
C_{\tau_n}=e^{-rh}\left\{\mathbb{E}[\px[h]{x+t_n}\Pi_{\tau_{n+1}} \mid \mathcal{F}_{t_n}]+\mathbb{E}[(1-\px[h]{x+t_n})\max(F_{t_{n+1}},G) \mid \mathcal{F}_{t_n}]\right\}.
\]
The expected payoff at time $t_n$ can be expressed as
\begin{align*}
\mathbb{E}[\Pi_{\tau_n^{\widehat{p}, K}}\mid \mathcal{F}_{t_n}] &= g_{t_n}\mathbf{1}_{A_n(\widehat{p}, K)}+C_{\tau_n^{\widehat{p}, K}}\mathbf{1}_{A_n(\widehat{p}, K)^c},\\
    \mathbb{E}[\Pi_{\tau_n}\mid \mathcal{F}_{t_n}] &= g_{t_n}\mathbf{1}_{A_n}+C_{\tau_n}\mathbf{1}_{A_n^c}\\
    &=g_{t_n}\mathbf{1}_{A_n}+C_{\tau_n}\mathbf{1}_{A_n^c}
    +C_{\tau_n}(\mathbf{1}_{A_n(\widehat{p}, K)^c}-\mathbf{1}_{A_n(\widehat{p}, K)^c}).
\end{align*}
Taking the difference of the two equations, we get
\begin{align*}
    &\mathbb{E}[\Pi_{\tau_n^{\widehat{p}, K}}-\Pi_{\tau_n}\mid \mathcal{F}_{t_n}]= g_{t_n}(\mathbf{1}_{A_n(\widehat{p}, K)}-\mathbf{1}_{A_n}) + \mathbb{E}[\px[h]{x+t_n}(\Pi_{\tau_{n+1}^{\widehat{p}, K}}-\Pi_{\tau_{n+1}})\mid \mathcal{F}_{t_n}]e^{-rh}\mathbf{1}_{A_n(\widehat{p}, K)^c}\\
    &+C_{\tau_n} (\mathbf{1}_{A_n(\widehat{p}, K)^c}-\mathbf{1}_{A_n^c})\\
    &=\left\{g_{t_n}-C_{\tau_n} \right\}(\mathbf{1}_{A_n(\widehat{p}, K)}-\mathbf{1}_{A_n})+\mathbb{E}[\px[h]{x+t_n}(\Pi_{\tau_{n+1}^{\widehat{p}, K}}-\Pi_{\tau_{n+1}})\mid \mathcal{F}_{t_n}]e^{-rh}\mathbf{1}_{A_n(\widehat{p}, K)^c}.
\end{align*}
\emph{Step 3 (Error decomposition).}\\
By the induction hypothesis, $\mathbb{E}[\Pi_{\tau_{n+1}^{\widehat{p}, K}}-\Pi_{\tau_{n+1}}|\mathcal{F}_{t_n}]\to 0$ in $L^2$ as $\widehat{p}, K\to \infty$. Hence, to prove $\mathbb{E}
[\Pi_{\tau_n^{\widehat{p}, K}}-\Pi_{\tau_n}\mid \mathcal{F}_{t_n}]\to 0$ in $L^2$ as $\widehat{p}, K\to \infty$, we need to prove
\[
L_n^{\widehat{p}, K}:=\left\{g_{t_n}-C_{{\tau_n}} \right\}(\mathbf{1}_{A_n(\widehat{p}, K)}-\mathbf{1}_{A_n})\to 0 \text{ in } L^2 \text{ as } \widehat{p}, K \to \infty.
\]
Obviously, $L_n^{\widehat{p}, K}=0$ for the sets ${A_n(\widehat{p}, K)\cap A_n}$ and ${A_n(\widehat{p}, K)^c\cap A_n^c}$.\\
When $A_n(\widehat{p}, K)^c\cap A_n$ (approximate strategy continues, true strategy exercises), we have $\Phi_{\widehat{p}}(\phi_n^{\widehat{p}, K}, \mathbf{x}_{t_n}^K)>g_{t_n}\geq C_{\tau_n}$, and when $A_n(\widehat{p}, K)\cap A_n^c$ (approximate strategy exercises, true strategy continues), we have $C_{\tau_n}>g_{t_n}\geq \Phi_{\widehat{p}}(\phi_n^{\widehat{p}, K}, \mathbf{x}_{t_n}).$ Hence, combining both cases yields
\begin{align}
    ||L_n^{\widehat{p}, K}||_{L^2}&\leq ||\Phi_{\widehat{p}}(\phi_n^{\widehat{p}, K}, \mathbf{x}_{t_n}^K)-C_{\tau_n}||_{L^2}\label{prop43_inequality}\\
    &\leq ||\Phi_{\widehat{p}}(\phi_n^{\widehat{p}, K}, \mathbf{x}_{t_n}^K)-C_{\tau_n^{\widehat{p}, K}}||_{L^2}+||C_{\tau_n^{\widehat{p}, K}}-C_{\tau_n}||_{L^2}\notag\\
    & \leq ||\Phi_{\widehat{p}}(\phi_n^{*(\widehat{p},K)}, \mathbf{x}_{t_n}^K)-C_{\tau_n^{\widehat{p}, K}}||_{L^2}+||C_{\tau_n^{\widehat{p}, K}}-C_{\tau_n}||_{L^2}\notag\\
    & \leq ||\Phi_{\widehat{p}}(\phi_n^{*(\widehat{p},K)}, \mathbf{x}_{t_n}^K)-C_{\tau_n}||_{L^2}+ 2||C_{\tau_n^{\widehat{p}, K}}-C_{\tau_n}||_{L^2}\notag\\
    & \leq||\Phi_{\widehat{p}}(\phi_n^{*(\widehat{p},K)}, \mathbf{x}_{t_n}^K)-C_{\tau_n}||_{L^2}+ 2e^{-rh}||\mathbb{E}[\px[h]{x+t_n}(\Pi_{\tau_{n+1}^{\widehat{p}, K}}-\Pi_{\tau_{n+1}})|\mathcal{F}_{t_n}]||_{L^2},\notag
\end{align}
 where $\phi_n^{*(\widehat{p},K)}$ is the best set of neural network parameters that approximates the true continuation value
\[
\phi_n^{*(\widehat{p},K)}=\arg \min_{\phi\in\Theta_{\widehat{p}}} ||\Phi_{\widehat{p}}(\phi, \mathbf{x}^K_{t_n})-C_{\tau_n}||_{L^2}.
\]
The second and fourth inequalities are derived by applying the triangular inequality, and the third inequality holds because $\phi_n^{\widehat{p}, K}$ is the optimal neural network parameters for $C_{\tau^{\widehat{p}, K}_n}$, but $\phi_n^*$ is not. Finally, the first term of the inequality converges to 0 according to the approximation result in Lemma \ref{lemma: approx}, and the second term converges to 0 by the induction hypothesis.
\end{proof}
\label{prop: k_m_converge}
\end{proposition}

Proposition \ref{prop: k_m_converge} only assures that the policy value obtained by the deep‑signature LSMC approach converges to the true value under the discrete time sense. We therefore need an additional result to analyze the Monte Carlo convergence. That is, the numerical value converges to the true value when (i) the time grid becomes finer and (ii) the number of Monte Carlo paths increases. Let $\widehat{X}^N=(\widehat{X}^{N,1}, \widehat{X}^{N,2}, \widehat{X}^{N,3})$ denote the discretized path of $\widehat{X}=(t, V_t, \mu_t)$ on a uniform grid with $N$ intervals $0=t_0<t_1<\cdots <t_N=T$.  $\text{Sig}^{\leq \infty}(\widehat{X}^N)_{[0,t]}$ is the signature of the discretized path. We make the following assumption similar to \cite{BayerPelizzariSchoenmakers2025} for every signature level $k\geq0$ and each index $(i_1, \cdots, i_k) \in \{1, \cdots,d\}$
\begin{equation}
\int_{0<u_1<\cdots<u_k<t} d\widehat{X}_{u_1}^{N,i_1}\cdots d\widehat{X}_{u_k}^{N,i_k}\xrightarrow[N \to \infty]{L^2}\int_{0<u_1<\cdots<u_k<t} d\widehat{X}_{u_1}^{i_1}\cdots d\widehat{X}_{u_k}^{i_k},
\label{eq: assump_discrete_sig_converge}
\end{equation}
which means the corresponding signature of the discretized path converges to that of the continuous path as the time step $T/N$ tends to 0. This is a direct result, as the rough path signature is defined as a continuous limit of Lipschitz paths (see \eqref{signature_smooth_approx} and arguments below it).

We let $\tau_n^{\widehat{p}, K, N}$ denote the optimal stopping rule under the NN's width $\widehat{p}$, truncated signature level $K$, and $N$ discretization steps. Furthermore, let $\tau_n^{\widehat{p}, K, N, (i)}$ be the realization of $\tau_n^{\widehat{p}, K, N}$ for the path $i$. The approximating stopping times $\tau_n^{\widehat{p}, K, N}$ and its realization $\tau_n^{\widehat{p}, K, N, (i)}$ are obtained by the dynamic programming principle
\begin{align*}
\tau_N^{\widehat{p}, K, N, (i)}&=t_N, \\    
\tau_n^{\widehat{p}, K, N, (i)}&=t_n\mathbf{1}_{\{g_{t_n}^{(i)}\geq \Phi_{\widehat{p}}(\phi_n^{\widehat{p}, K, N, M}, \mathbf{x}_{t_n}^{k,N,(i)})\}}+\tau_{n+1}^{\widehat{p}, K, N, (i)}\mathbf{1}_{\{g_{t_n}^{(i)}<\Phi_{\widehat{p}}(\phi_n^{\widehat{p}, K, N, M}, \mathbf{x}_{t_n}^{k,N,(i)})\}}, \\
\tau_N^{\widehat{p}, K, N}&=t_N, \\    
\tau_n^{\widehat{p}, K, N}&=t_n\mathbf{1}_{\{g_{t_n}\geq \Phi_{\widehat{p}}(\phi_n^{\widehat{p}, K, N}, \mathbf{x}_{t_n}^{k,N})\}}+\tau_{n+1}^{\widehat{p}, K, N}\mathbf{1}_{\{g_{t_n}<\Phi_{\widehat{p}}(\phi_n^{\widehat{p}, K, N}, \mathbf{x}_{t_n}^{k,N})\}}.
\end{align*} 

Subsequently, we can define $\phi_n^{\widehat{p}, K, N}$ as the optimal neural network that solves the optimization problem
\begin{equation}
\phi_n^{\widehat{p}, K, N}=\arg \min_{\phi\in\Theta_{\widehat{p}}} ||\Phi_{\widehat{p}}(\phi, \mathbf{x}^{K,N}_{t_n})-C_{\tau_n^{\widehat{p}, K, N}}||_{L^2},
\label{eq: opt_pKN}
\end{equation}
and further define $\phi_n^{\widehat{p}, K, N, M}$ as the optimal neural network solves the sample average approximation of the optimization problem \eqref{eq: opt_pKN} with $M$ sample paths
\[
\phi_n^{\widehat{p}, K, N, M}=\arg \min_{\phi_n\in\Theta_{\widehat{p}}} \dfrac{1}{M}\sum_{i=1}^M\left(\Phi_{\widehat{p}}(\phi_n, \mathbf{x}_{t_n}^{K,N,(i)})-C_{\tau_n^{\widehat{p}, K, N, (i)}}^{(i)}\right)^2.
\]

We follow the notation in \cite{ClementLambertonProtter2002} and \cite{LapeyreLelong2021} to define parameters $\phi:=(\phi_1, \cdots, \phi_{N-1})\in \Theta_{\widehat{p}}^{N-1}$, vectors $(\mathbf{x}_1, \cdots, \mathbf{x}_N)\in (\mathbb{R}^d)^N$, $(F_1^{VA}, \cdots,F_N^{VA})\in \mathbb{R}^N,$ $\left((\px[h]{x}, \px[2h]{x}, \cdots, \px[Nh]{x}),(\px[h]{x+t_1},  \px[2h]{x+t_1}, \cdots, \px[(N-1)h]{x+t_1}), \cdots, (\px[h]{x+t_{N-1}})\right)\in \prod_{k=0}^{N-1} \mathbb{R}^{N-k}$. We furthre define a vector field $F^{VA}=(F^{VA}_1, \cdots,F_N^{VA})$ by 
\begin{align*}
F^{VA}_N(\phi, \mathbf{x}_N) &= g_N, \\
F^{VA}_n(\phi, \mathbf{x}_n) &= g_{t_n}\mathbf{1}_{\{g_{n+1}\geq\Phi_{\widehat{p}}(\phi_{n+1}, \mathbf{x}_{n+1})\}}\\
&+[\px[h]{x+t_n}F^{VA}_{n+1}(\phi, \mathbf{x}_{n+1})
+(1-\px[h]{x+t_n})\max({G, F_{n+1}})]e^{-rh}\mathbf{1}_{\{g_{n+1}<\Phi_{\widehat{p}}(\phi_{n+1}, \mathbf{x}_{n+1})\}}.
\end{align*}

After the preparations above, we still need two more lemmas before obtaining the final convergence result. The first lemma is Lemma \ref{lemma: abs_diff_F}, which gives an upper bound of the difference in VA prices obtained by two different sets of neural network parameters $\phi^a, \phi^b$.
\begin{lemma}\label{lemma: abs_diff_F}
For $j=1 \text{ to }N-1,$ we have
\begin{align*}
&\left|F^{VA}_n(\phi^a, \mathbf{x}_n)-F^{VA}_n(\phi^b, \mathbf{x}_n)\right|\\
&\leq  \left(\sum_{j=n}^{N} e^{-r(j-n)h}\px[(j-n)h]{x+t_n} g_{t_j} + \sum_{j=n+1}^N e^{-r(j-n)h}(1-\px[(j-n)h]{x+t_n})\max(F_{j+1},G)\right)\\
    &\times \sum_{j=n}^{N-1} \left|\mathbf{1}_{\{|g_{t_n}-\Phi_{\widehat{p}}(\phi^a_n, \mathbf{x}_n)| \leq |\Phi_{\widehat{p}}(\phi^a_n, \mathbf{x}_n)-\Phi_{\widehat{p}}(\phi^b_n, \mathbf{x}_n)|\}}\right|.    
\end{align*}

\begin{proof}
    Let $A_n=\{g_{t_n}\geq \Phi_{\widehat{p}}(\phi_n^a, \mathbf{x}_n)\}$, $\widetilde{A}_n=\{g_{t_n}\geq \Phi_{\widehat{p}}(\phi_n^b, \mathbf{x}_{t_n})\}.$ We have 
    \begin{align*}
    &F^{VA}_n(\phi^a, \mathbf{x}_n)-F^{VA}_n(\phi^b, \mathbf{x}_n) = g_{t_n}(\mathbf{1}_{A_n}-\mathbf{1}_{\widetilde{A}_n})
\\
&+\sum_{j=n+1}^{N-1}e^{-r(j-n)h}\px[(j-n)h]{x+t_n}g_{t_j}(\mathbf{1}_{A^c_n\cdots A_{j-1}^cA_j}-\mathbf{1}_{\widetilde{A}^c_n\cdots \widetilde{A}^c_{j-1}\widetilde{A}_j})\\
&+ e^{-r(N-n)h}\px[(N-n)]{x+t_n}g_N(\mathbf{1}_{A^c_n\cdots A_{N-1}^c}-\mathbf{1}_{\widetilde{A}^c_n\cdots \widetilde{A}^c_{N-1}})\\
&+\sum_{j=n+1}^Ne^{-r(j-n)h}(1-\px[(j-n)h]{x+t_n})\max(F_{j+1},G)(\mathbf{1}_{A^c_n\cdots A_{j-1}^c}-\mathbf{1}_{\widetilde{A}^c_n\cdots \widetilde{A}^c_{j-1}}).
    \end{align*}
In addition, we have the following inequalities
\begin{align}
    |\mathbf{1}_{A_n}-\mathbf{1}_{\widetilde{A}_n}| &= \mathbf{1}_{\{\Phi_{\widehat{p}}(\phi^a_n, \mathbf{x}_n)\leq g_{t_n} < \Phi_{\widehat{p}}(\phi^b_n, \mathbf{x}_n)\}} +\mathbf{1}_{\{\Phi_{\widehat{p}}(\phi^b_n, \mathbf{x}_n)\leq g_{t_n} < \Phi_{\widehat{p}}(\phi^a_n, \mathbf{x}_n)\}}\notag\\
     &\leq \mathbf{1}_{\{|g_{t_n}-\Phi_{\widehat{p}}(\phi^a_n, \mathbf{x}_n)| \leq |\Phi_{\widehat{p}}(\phi^a_n, \mathbf{x}_n)-\Phi_{\widehat{p}}(\phi^b_n, \mathbf{x}_n)|\}},
    \label{ineq: A}\\
    |\mathbf{1}_{A^c_n\cdots A_{j-1}^cA_j}-\mathbf{1}_{\widetilde{A}^c_n\cdots \widetilde{A}^c_{j-1}\widetilde{A}_j}| &\leq \sum_{k=n}^{j-1} |\mathbf{1}_{A^c_k}-\mathbf{1}_{\widetilde{A}_k^c}| + |\mathbf{1}_{{A_j}}-\mathbf{1}_{\widetilde{A}_j}|=\sum_{k=n}^j |\mathbf{1}_{A^c_k}-\mathbf{1}_{\widetilde{A}^c_k}|, \notag\\
    |\mathbf{1}_{A^c_n\cdots A_{j-1}^c}-\mathbf{1}_{\widetilde{A}^c_n\cdots \widetilde{A}^c_{j-1}}| &\leq \sum_{k=n}^{j-1} |\mathbf{1}_{A^c_k}-\mathbf{1}_{\widetilde{A}^c_k}|.\notag
\end{align}
Therefore, we have
\begin{align*}
    &\left|F^{VA}_n(\phi^a, \mathbf{x}_n)-F^{VA}_n(\phi^b, \mathbf{x}_n)\right| \\&\leq  \left(\sum_{j=n}^{N} e^{-r(j-n)h}\px[(j-n)h]{x+t_n} g_{t_j} + \sum_{j=n+1}^N e^{-r(j-n)h}(1-\px[(j-n)h]{x+t_n})\max(F_{j+1},G)\right)\\
    &\times \sum_{j=n}^{N-1} |\mathbf{1}_{A_j}-\mathbf{1}_{\widetilde{A}_j}|.
\end{align*}
Combining the result and (\ref{ineq: A}), we obtain the result of the lemma.
\end{proof}
\label{lemma4}
\end{lemma}

Following \cite{LapeyreLelong2021}, we make the following four assumptions for the regularity of the neural network function.
\begin{assumption}\label{ass:A1}
For every $\widehat{p}\in\mathbb{N}$, $\widehat{p}>1$, there exist $q\ge 1$ and $\kappa_{\widehat{p}}>0$ such that
\[
    \forall x\in\mathbb{R}^{d_{K}+1}, \ \forall \phi\in\Theta_{\widehat{p}}, \qquad
    \bigl|\Phi_{\widehat{p}}(\phi, \mathbf{x})\bigr|
    \le \kappa_{\widehat{p}}\bigl(1+|x|^q\bigr),
\]
where $d_{K}+1$ is the dimension of the time-augmented path's signature truncated at level $K$. Moreover, for all $1\le n\le N-1$, a.s. the random functions
\[
    \phi \in \Theta_{\widehat{p}} \longmapsto \Phi(\phi, \mathbf{x}_{t_n})
\]
are continuous. Since $\Theta_{\widehat{p}}$ is a compact set, the continuity automatically yields the uniform continuity.
\end{assumption}
\begin{assumption}\label{ass:A2}
For $q$ defined in Assumption~\ref{ass:A1}, 
\[
    \mathbb{E}\bigl[ |\mathbf{x}_{t_n}|^{2q}\bigr] < \infty
    \quad\text{for all } 0\le n\le N .
\]
\end{assumption}

\begin{assumption}\label{ass:A3}
For all $\widehat{p}\in\mathbb{N}$, $\widehat{p}>1$, and all $1\le n\le N-1$,
\[
    \mathbb{P}\bigl( \Pi_{\tau_n} = \Phi_{\widehat{p}}(\phi_n^{\widehat{p}}, \mathbf{x}_{t_n}) \bigr) = 0 .
\]
We denote the set of neural network parameters that minimize the square loss
\[
\mathcal{S}_n^{\widehat{p}}=\arg  \inf_{\phi \in \Theta_{\widehat{p}}} \mathbb{E}\left[|\Phi_{\widehat{p}}(\phi, \mathbf{x}_{t_n})-C_{\tau_n}|^2\right].
\]
\end{assumption}

\begin{assumption}\label{ass:A4}
For every $\widehat{p}\in\mathbb{N}$, $\widehat{p}>1$, and every $1\le n\le N$, for all
$\phi^a, \phi^b\in S_n^{\widehat{p}}$,
\[
    \Phi_{\widehat{p}}(\phi^a,\mathbf{x}) = \Phi_{\widehat{p}}(\phi^b,\mathbf{x})
    \qquad\text{for all } \mathbf{x}\in\mathbb{R}^{d_{K}+1}.
\]
\end{assumption}

With all the assumptions above, we can arrive at the second lemma, which shows the convergence of the neural network approximation as the path number goes to infinity. 
\begin{lemma}\label{lemma9}
Let $\phi_n^{\widehat{p}, K}\in\mathcal{S}_n^{\widehat{p}}$ be the optimal neural network parameters and $\phi_n^{\widehat{p}, K,M}$ be the minimizer of the sample average loss
\[
\phi^{\widehat{p}, K,M}_n:=\arg \inf_{\phi\in\Theta_{\widehat{p}}} \dfrac{1}{M}\sum_{i=1}^M|\Phi_{\widehat{p}}(\phi, \mathbf{x}_{t_n}^{(i)})-C^{\widehat{p}, K, (i)}_{t_n}|^2.
\]
Subsequently, for $n=1, \cdots,N-1,$ $\Phi_{\widehat{p}}(\phi_n^{\widehat{p}, K,M}, \mathbf{x}_{t_n})\xrightarrow{a.s.}\Phi_{\widehat{p}}(\phi_n^{\widehat{p}, K}, \mathbf{x}_{t_n})$ as $M\rightarrow \infty$.
\begin{proof}
This is a direct result of Proposition 4.8 in \cite{LapeyreLelong2021} with their Lemma 4.9 replaced by our Lemma \ref{lemma4}.
\end{proof}
\end{lemma}

Lastly, with all the preparations in this section, we can finally prove that the VA price from the deep signature LSMC approach converges to the true price.
\begin{proposition}
Given Assumptions \ref{ass:A1}-\ref{ass:A4}, we have
\[
\lim_{M\rightarrow\infty, \widehat{p}\rightarrow\infty,K\rightarrow\infty, N\rightarrow\infty}\dfrac{1}{M}\sum_{i=1}^M\Pi^{(i)}_{\tau_n^{\widehat{p}, K, N, (i)}} = \mathbb{E}[\Pi_{\tau_n}].
\]
\begin{proof}
First, we can rewrite the inequality \eqref{prop43_inequality} to the following form by the triangular inequality
\begin{align*}
||L_n^{\widehat{p}, K, N}||_{L^2} &\leq || \Phi_{\widehat{p}}(\phi_n^{\widehat{p}, K, N}, \mathbf{x}_{t_n})-C_{\tau_n} ||_{L^2}\\ 
&\leq || \Phi_{\widehat{p}}(\phi_n^{\widehat{p}, K}, \mathbf{x}_{t_n})-C_{\tau_n} ||_{L^2} + || \Phi_{\widehat{p}}(\phi_n^{\widehat{p}, K, N}, \mathbf{x}_{t_n})-\Phi_{\widehat{p}}(\phi_n^{\widehat{p}, K}, \mathbf{x}_{t_n})||_{L^2}.
\end{align*}
From Proposition \ref{prop: k_m_converge}, the first term converges to $0$. The second term also converges to $0$ by the assumption in (\ref{eq: assump_discrete_sig_converge}) and the universal approximation theorem. Therefore, we have 
\[
\lim_{ \widehat{p}\rightarrow\infty, K\rightarrow\infty, N\rightarrow\infty}\mathbb{E}[\Pi_{\tau^{\widehat{p}, K, N}_n}\mid \mathcal{F}_{t_n}] = \mathbb{E}[\Pi_{\tau_n}\mid \mathcal{F}_{t_n}] \text{ in } L^2.
\]
Next, we need to prove that 
\[
\lim_{M\rightarrow\infty}\dfrac{1}{M}\sum_{i=1}^M\Pi^{(i)}_{\tau_n^{\widehat{p}, K, N, (i)}} = \mathbb{E}[\Pi_{\tau_n^{\widehat{p}, K, N}}]. 
\]
By strong law of large numbers,
\[
\lim_{M\rightarrow\infty}\dfrac{1}{M}\sum_{i=1}^MF^{VA}_n(\phi, \mathbf{x}_{t_n}^{(i)})\xrightarrow{a.s.}\mathbb{E}[F^{VA}_n(\phi, \mathbf{x}_{t_n})] .
\]
Therefore, it suffices to prove
\[
\Delta F^{VA}_M:=\dfrac{1}{M}\sum_{i=1}^M(F^{VA}_n(\phi^{\widehat{p}, K,M}, \mathbf{x}_{t_n}^{(i)})-F^{VA}_n(\phi^{\widehat{p}, K}, \mathbf{x}_{t_n}^{(i)}))\xrightarrow[M\rightarrow\infty]{a.s.}0.
\]
From Lemma \ref{lemma4}, we have
\begin{align*}
|\Delta F^{VA}_M| &\leq\dfrac{1}{M}\sum_{i=1}^M\left|F^{VA}_n(\phi^{\widehat{p}, K,M}, \mathbf{x}_{t_n}^{(i)})-F^{VA}_n(\phi^{\widehat{p}, K}, \mathbf{x}_{t_n}^{(i)})\right|\\
&\leq\dfrac{1}{M}\sum_{i=1}^M\left(\sum_{j=n}^N\max_{n\leq k\leq N}\left|(e^{-r(k-n)h}\px[(k-n)h]{x+t_n}^{(i)} g_{t_j}^{(i)} \right| \right. \\
&\left. \qquad
+ \sum_{j=n+1}^N\max_{n\leq k \leq N}\left| e^{-r(k-n)h}(1-\px[(k-n)h]{x+t_n}^{(i)})\max(G, F_{j+1}^{(i)})\right|\right)\\
&\times\left( \sum_{j=n}^{N-1}\mathbf{1}_{\left\{\left|g_{t_j}^{(i)}-\Phi_{\widehat{p}}(\phi^{\widehat{p}, K,M}_j, \mathbf{x}_{t_j}^{(i)})\right| \leq \left|\Phi_{\widehat{p}}(\phi^{\widehat{p}, K,M}_j, \mathbf{x}_{t_j}^{(i)})-\Phi_{\widehat{p}}(\phi^{\widehat{p}, K}_j, \mathbf{x}_{t_j}^{(i)})\right|\right\}}\right).
\end{align*}
By Lemma \ref{lemma9}, $\left|\Phi_{\widehat{p}}(\phi_n^{\widehat{p}, K,M}, \mathbf{x}_{t_n}^{(i)})-\Phi_{\widehat{p}}(\phi_n^{\widehat{p}, K}, \mathbf{x}_{t_n}^{(i)})\right|\rightarrow0$ for all $i=1,\cdots,M$ and $n=1,\cdots,N-1$ as $M\rightarrow\infty.$ Finally, for any $\varepsilon>0$,
\begin{align*}
&\limsup_M|\Delta F^{VA}_M| \\
\leq& \limsup_M
    \dfrac{1}{M}\sum_{i=1}^M\left(\sum_{j=n}^N\max_{n\leq k\leq N}\left|e^{-r(k-n)h}\px[(k-n)h]{x+t_n}^{(i)} g_{t_j}^{(i)} \right| \right. \\
    & \left. + \sum_{j=n+1}^N\max_{n\leq k \leq N} \left| e^{-r(k-n)h}(1-\px[(k-n)h]{x+t_n}^{(i)})\max(G, F_{j+1}^{(i)})\right|\right)
\times \sum_{j=n}^{N-1}\left|\mathbf{1}_{\left\{\left|g_{t_j}^{(i)}-\Phi_{\widehat{p}}(\phi^{\widehat{p}, K,M}_j, \mathbf{x}_{t_j}^{(i)})\right| \leq \varepsilon \right\}}\right|\\
= &\mathbb{E}\left[\left(\sum_{j=n}^N\max_{n\leq k\leq N}\left|e^{-r(k-n)h}\px[(k-n)h]{x+t_n} g_{t_j}\right| \right.\right.\\
&\left.\left.+ \sum_{j=n+1}^N\max_{n\leq k \leq N}\left| e^{-r(k-n)h}(1-\px[(k-n)h]{x+t_n})\max(G, F_{j+1})\right|\right)\times \sum_{j=n}^{N-1}\left|\mathbf{1}_{\left\{\left|g_{t_j}-\Phi_{\widehat{p}}(\phi^{\widehat{p}, K,M}_j, \mathbf{x}_{t_j})\right| \leq \varepsilon\right\}}\right|\right],
\end{align*}
where the last equality follows from the law of large numbers. Letting $\varepsilon\rightarrow0$ and by Assumption \ref{ass:A3}, we prove that $\limsup |\Delta F^{VA}_M|\rightarrow0$  .
\end{proof}
\end{proposition}

\section{Numerical results}
\label{sec: numerical}
In this section, we present numerical results for the valuation of a variable annuity contract with an early-surrender option under a non-Markovian framework. The financial market model and the stochastic mortality model are assumed to follow the dynamics in Section \ref{sec: problem_statement}. We first summarize the baseline parameters for the contract setting, and then conduct a series of sensitivity analyses. The numerical results are obtained using the deep signature LSMC method described in Section \ref{sec: Optimal_surrender}.

\subsection{Model and VA contract specification}
We assume the baseline contract is a 20-year variable annuity ($T=20$) issued to a policyholder aged 60. The initial premium is set to $100$ and is fully invested in the VA account (i.e., $F_0=100$). Moreover, the contract provides two types of guarantee, GMMB and GMDB, each with a guarantee amount of $G=100.$ Policyholders are allowed to surrender the policy prior to maturity. Upon surrender at time $t<T$, the policyholder receives the surrender payoff
\[
g_t=(1-\kappa_t)F_t, \text{ where } \kappa_t=1-e^{-\kappa(T-t)}.
\]
Hence, the surrender payoff is $e^{-\kappa(T-t)}F_t$. In the baseline case, we set $\kappa=0.2\%.$ The underlying equity is modeled by the rough Heston model, as shown in (\ref{eq: heston}). 
The model parameters are obtained from the calibration results in \cite{Jeng2021SPX}, which are calibrated to options price data of the S\&P 500 index with third-order 
$\mathrm{Pad\acute{e}}$ approximant ($\mathrm{GPad\acute{e}}$).
The parameters in the Volterra mortality model are calibrated using the method mentioned in Section \ref{sec: mortality_model}. All the baseline parameters above are summarized in Table \ref{tab:Param_base}. Note that the initial force of mortality $\mu_x$, as shown in Table \ref{tab:Param_base}, is different from that used in calibration (Table \ref{table:params_mort_cali}). This discrepancy arises because we apply mortality data for ages 30 to 80 in the calibration process, while we only consider policyholders aged 60 to 80 in variable annuity pricing.
\begin{table}[htbp]
    \caption{Parameters for the base scenario}
    \centering 
    \begin{tabular}{crcrcr}   
    \toprule
    Parameter & Value & Parameter & Value & Parameter & Value \\
    \midrule
    \multicolumn{6}{l}{Variable annuity policy} \\ 
    $T$ & $20$ & $x$ & $60$  &$F_0$ & $100$ \\
    $G$ & $100$ &  $\kappa$ & $0.002$\\
    \midrule
    \multicolumn{6}{l}{Financial market} \\
    $S_0$ & $100$ & $V_0$ & $0.2720$ & $\gamma$ & $0.1206$\\
     $\theta$ & $0.0721$ & $\nu$ & $0.2897$ & $\rho$ & $-0.7445$\\
     $H_S$ & $0.0286$  & $r$ & $0.0400$\\
     \midrule
    \multicolumn{6}{l}{Mortality model} \\
    $\mu_x$ & $0.018999$  & $\lambda$ &$0.047780$ & $\sigma$ & $0.005023$\\
     $H_m$ & $0.703932$ \\
    \bottomrule
    \end{tabular}
     \label{tab:Param_base}
\end{table}

\subsection{Numerical implementation}

We use a monthly time step for the underlying simulation $(\Delta t = 1/12)$, allowing early surrender at discrete monthly time points. For the deep signature LSMC algorithm, we simulate $M=2^{17}$ sample paths for training the neural network models, and another $\widetilde{M}=2^{17}$ independent test paths for calculating the estimated VA policy value, of the stochastic process $(S_t, V_t, \mu_{x+t})$. The stock price, volatility, and the force of mortality paths $(S_t, V_t, \mu_{x+t})$ are simulated numerically according to the method in Section \ref{sec: Forward simulation of the underlying process}. Furthermore, the realized survival probabilities are calculated as \eqref{eq: survival_prob} to compute the monthly decrements along each path.

We approximate the continuation value using a deep neural network with signature features, as described in Section \ref{sec: State Variable Encoding and Neural Network Structure}.  For each exercise date, the network is trained by minimizing the mean squared error between the realized continuation payoffs and the neural network predictions using all the paths. During the training, the neural network parameters are initialized using the weights and biases from the previous time step to save training time and stabilize the results. 

We employ early stopping with an ``80\%-20\%" validation split to mitigate overfitting. A two-pass procedure is used to remove look-ahead bias: the first set of paths (training set) is used to learn the surrender strategy, and the second set of paths (test set), which is independent of the training paths, is used to evaluate the VA policy with the learned strategy. The reported prices and fair fees in this section are computed from the test set.

\subsection{Fair fee and sensitivity analysis}
The VA price is a decreasing function of the fee rate $c$, allowing us to solve the fair fee $c^*$ numerically by a bisection algorithm. For a fixed fee $c$, we run the deep signature LSMC algorithm to obtain the corresponding VA price $U_0(c)$, and then update the fee interval $[c^L, c^U]$ according to whether the price is above or below the initial premium $F_0$. The bisection procedure is terminated when the interval length $c^U-c^L$ is below a tolerance level. In the numerical examples, we set this level to $10^ {-4}$.

To understand the effects of different parameters on the fair fee, we conduct sensitivity analysis for $H_S$, the Hurst parameter in the rough Heston model, and $H_m$, the Hurst parameter in the stochastic mortality model. The resulting fair fees are reported in Table \ref{tab: sensi_fee}. We also compute the fair fees when early surrender is not allowed and find that they are all lower than those when early surrender is allowed, indicating additional policy value from the early surrender feature.

\begin{table}[h!]
\centering
\caption{Sensitivity analysis of VA fair fee $c^*$ on the Hurst parameters $H_S$ and $H_m$}
\begin{tabular}{cccccc}
\toprule
$H_S$ & $0.05$ & $0.15$ & $0.25$ & $0.35$ & $0.45$ \\
\midrule
 $c^*$ with & $0.952\%$  & $0.966\%$ & $0.989\%$ & $1.020\%$  & $1.039\%$ \\
\midrule
 $c^*$ without & $0.724\%$ & $0.746\%$ & $0.767\%$ & $0.785\%$ & $0.800\%$\\
\midrule
$H_m$ & $0.55$ & $0.65$ & $0.75$ & $0.85$ & $0.95$ \\
\midrule
$c^*$ with & $0.922\%$  & $0.941\%$ & $0.949\%$ & $0.993\%$ & $1.021\%$\\
\midrule
$c^*$ without & $0.693\%$ & $0.710\%$& $0.728\%$& $0.751\%$ & $0.777\%$  \\
\bottomrule
\end{tabular}
\label{tab: sensi_fee}
\begin{flushleft}
 Average Computation time for each $c^*$ with early surrender: 3.5 Hours. ``$c^*$ with'' means the fair fee with early surrender. ``$c^*$ without'' means the fair fee without early surrender. 
\end{flushleft}
\end{table}

We analyze the impact of the Hurst parameter $H_S$ in the rough Heston model on the fair fee. We observe that the fair fee $c^*$ increases as $H_S$ increases, from $H_S=0.05$ to $H_S=0.45$. In particular, the Hurst parameter represents the ``roughness" of the volatility path. A small Hurst parameter makes the fractional kernel, $K(t-s)=(t-s)^{H_S-1/2}/\Gamma(H_S+1/2)$, decay more quickly in $(t-s)$ so that the contribution of the past volatility values is dampened. In other words, with a small Hurst parameter, the variance path exhibits shorter memory and behaves closer to a mean-reverting feature towards $\theta$, and hence the volatility tends to fluctuate around the long-term mean level. As a result, less extreme values are observed in the volatility paths and the distribution of fund values $F_t$ is less spread out. Therefore, the embedded benefit guarantees are less valuable when $H_S$ is small, making the fair fee $c^*$ lower. Furthermore, \cite{HanWong2021TimeInconsistencyRoughVol} also suggest that the effect of the Hurst parameter may differ between short and long time horizons. However, our problem concerns the valuation of variable annuities, which are long-term contracts. Therefore, we only need to consider long-horizon behavior, where we find that a smaller Hurst parameter in the rough Heston model yields a smaller VA value and, hence, lower fee rates.

\begin{figure}[h!]
    \centering
    \includegraphics[width=5.5in, height=3.5in]{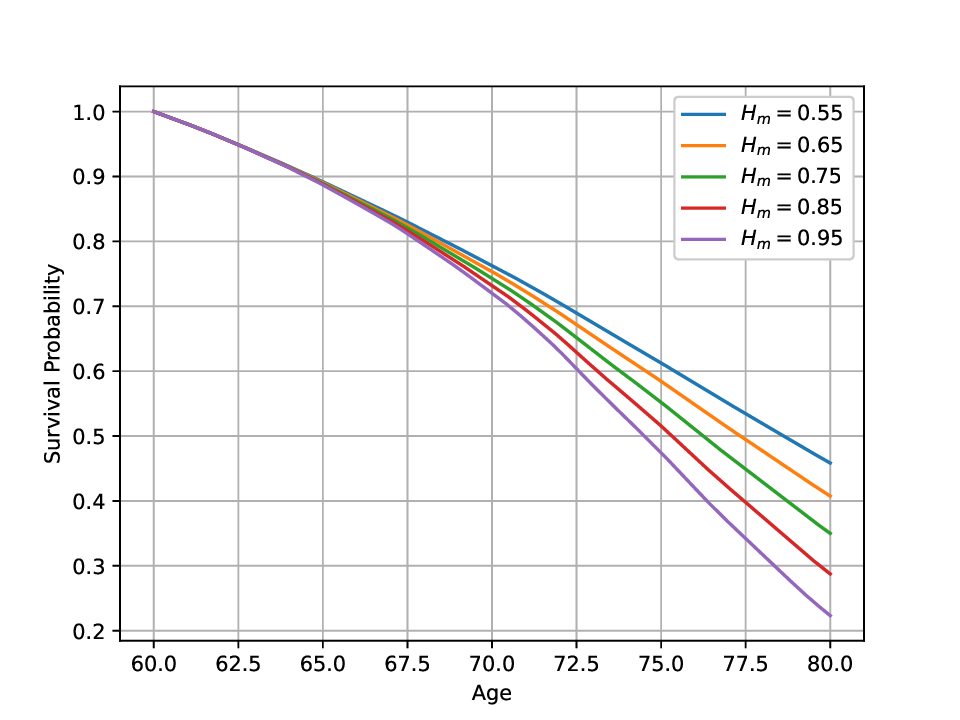}
    \caption{Sample survival probabilities with different $H_m$}
    \label{fig: sensi_Hm_survival_prob}
\end{figure}

\begin{figure}
    \centering
\includegraphics[width=5.5in, height=3.5in]{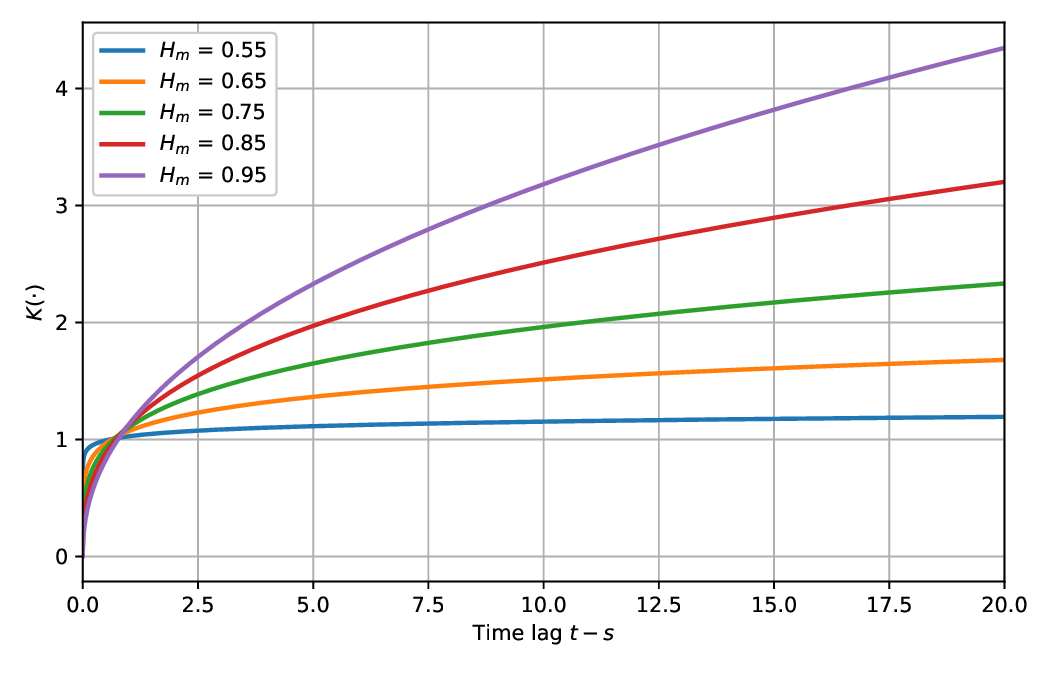}
    \caption{Kernel function with different Hurst parameter $H_m$}
    \label{fig: kernel}
\end{figure}

For the Hurst parameter in the mortality model, we first examine its effect on survival probabilities. Figure \ref{fig: sensi_Hm_survival_prob} shows how the mortality Hurst parameter $H_m$ affects the survival probability under the same Brownian motions. We observe systematically higher mortality intensities as $H_m$ increases, along with lower survival probabilities. This can be explained by the properties of the mortality kernel function and the long-term feature of the variable annuity. Figure  
\ref{fig: kernel} illustrates the kernel function value for $H_m=0.55, 0.65, 0.75, 0.85, 0.95$. Recall that the fractional kernel in the Volterra mortality model is $K(t-s)=(t-s)^{H_m-1/2}/\Gamma(H_m+1/2)$. Thus, a higher $H_m$ yields a slightly lower kernel value when the time lag $t-s$ is small, but the kernel value increases significantly as the time lag increases. Due to the special form of SVIE \eqref{eq: SVIE_mort}, $\mu_{x+t}$ accumulates more strongly over time when the mortality kernel function is high, causing a higher force of mortality in the long run. Ultimately, the monotonicity of survival probabilities helps explain the fair-fee pattern: the fair fee $c^*$ increases as $H_m$ increases. Specifically, when $H_m$ is larger and the survival probabilities are lower, the policyholder is more likely to receive the death benefit (with guarantee) early. This leads to a higher expected present value of the guarantee. Consequently, the insurer must charge a higher premium to compensate for the higher guaranteed benefits it provides.

\begin{figure}[htbp]
    \centering
    \begin{subfigure}{0.38\textwidth}
        \centering
        \includegraphics[width=1.2\textwidth,height=1.2\textwidth]{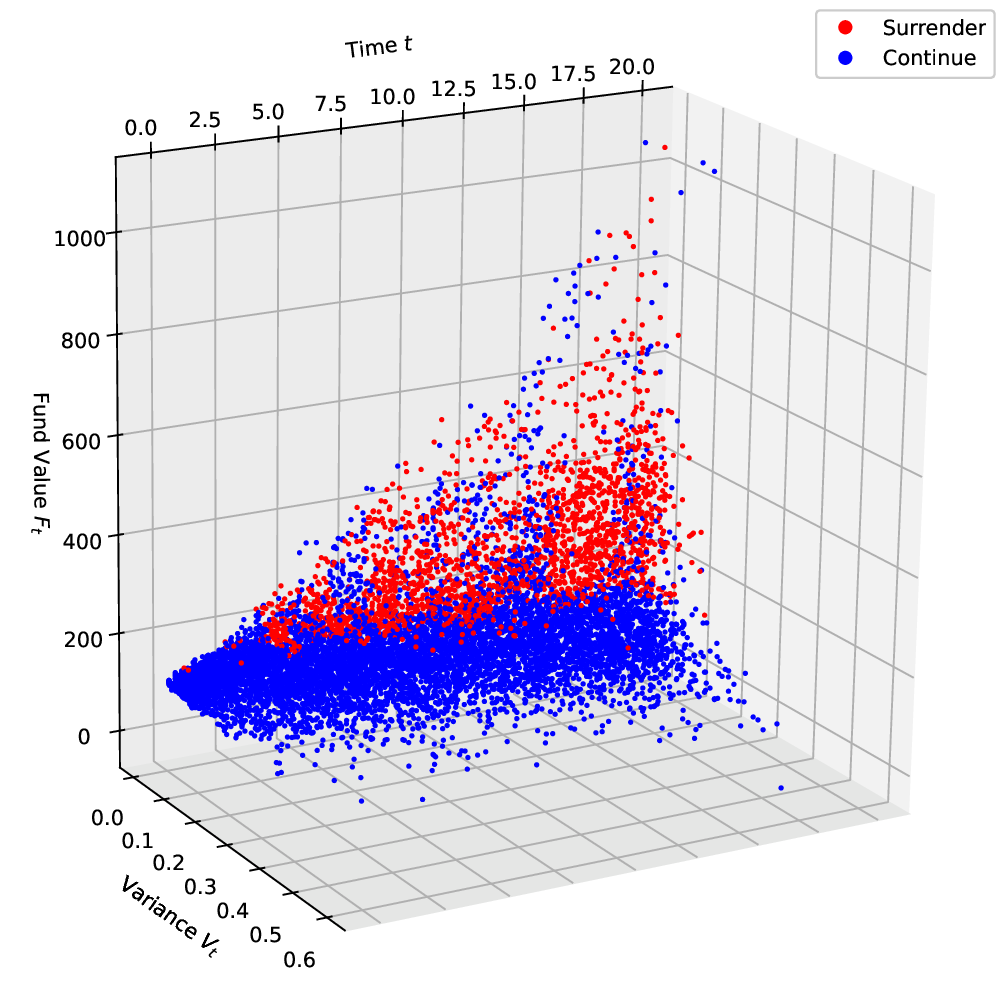}
        \caption{Surrender with \((t, F_t, V_t)\)}
    \end{subfigure}
    \hspace{1.3cm}
    \begin{subfigure}{0.38\textwidth}
        \centering
        \includegraphics[width=1.2\textwidth,height=1.2\textwidth]{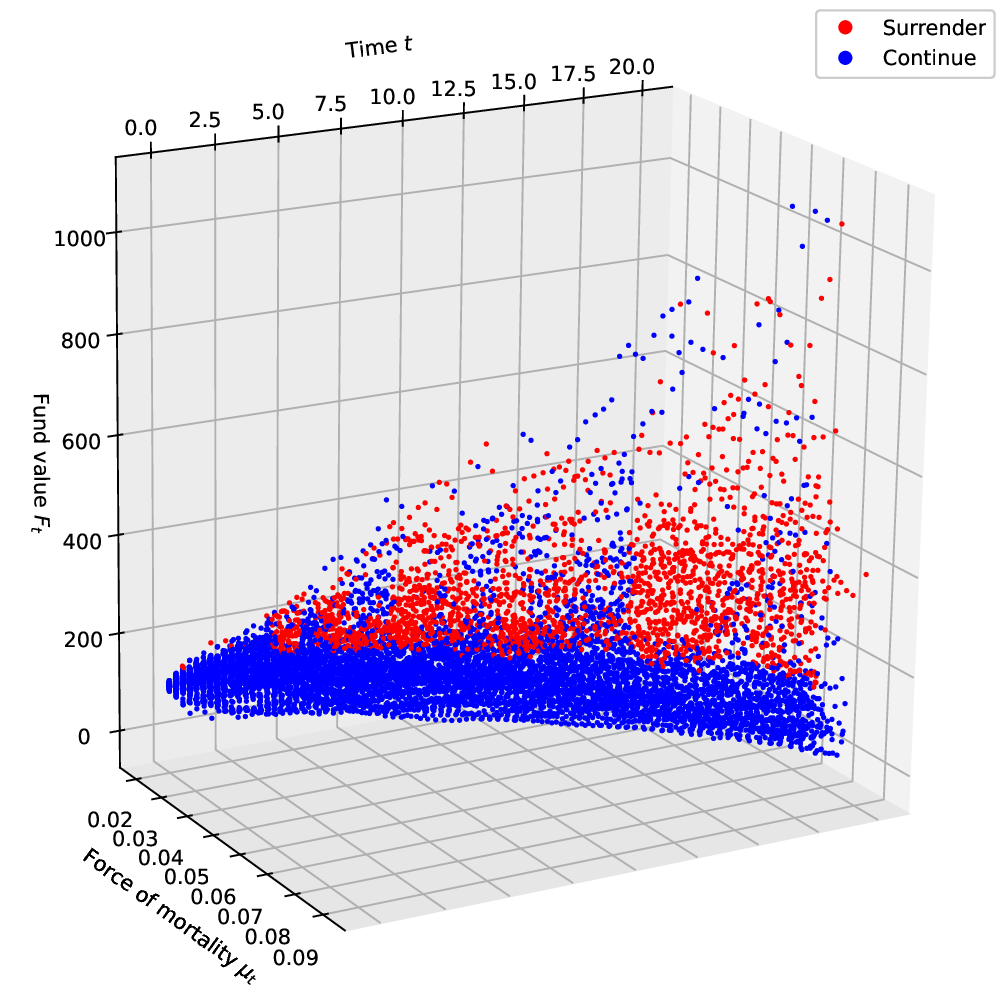}
        \caption{Surrender with \((t, F_t, \mu_{x+t})\)}
    \end{subfigure}
    \caption{3D surrender strategy for the baseline case.}
    \label{fig: boundary_base}
\end{figure}

\begin{figure}[!htbp]
    \centering
    \includegraphics[width=5.5in, height=3.0in]{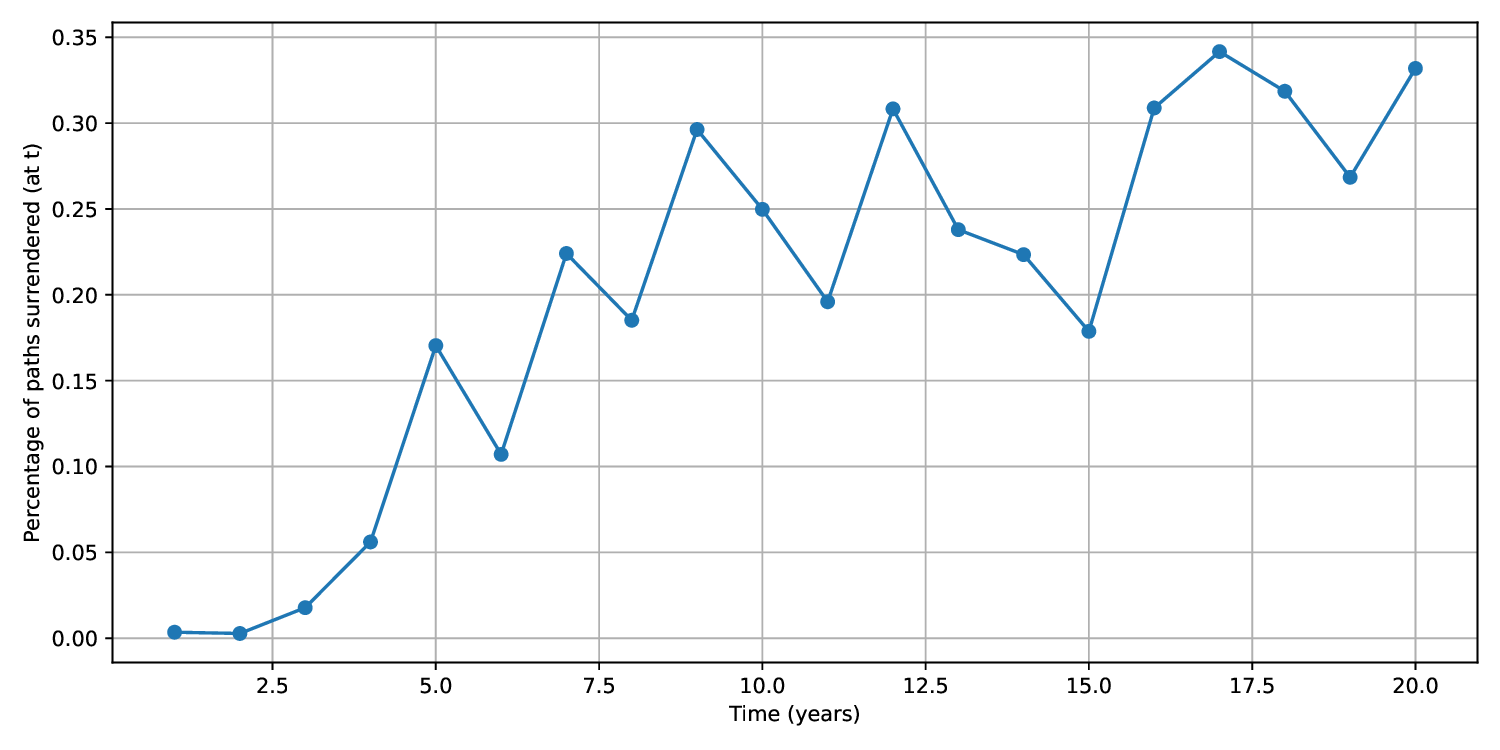}
    \caption{Surrender ratio of simulated paths over the years}
    \label{fig: surrender_rate_bench}
\end{figure}

We visualize the early surrender strategy of the baseline case in Figures \ref{fig: boundary_base} - \ref{fig: surrender_rate_bench}. However, due to the limitation of the visualization tool, we can only observe the ``non-Markovian" surrender behavior from a ``Markovian'' perspective. Specifically, Figure \ref{fig: boundary_base} presents two three-dimensional projections of the learned surrender policy: one in the space \((t, F_t, V_t)\) and the other in the space \((t, F_t, \mu_{x+t})\). Each point represents a simulated path at a surrender decision time. The points are colored according to the learned surrender strategy: red indicates surrender, and blue indicates continuation. From Figure \ref{fig: boundary_base}, we observe that more paths are exercised when the fund value $F_t$ is high and both $V_t$ and $\mu_{x+t}$ are at a low level. Although Figure \ref{fig: boundary_base} provides a tendency of surrender with state variables, we don't see a clear surrender boundary between surrender and continuation regions. This is because we observe the optimal decision from a low-dimensional Markovian perspective, using the current state variables $(t, F_t, V_t, \mu_{x+t})$ and ignoring the model's non-Markovian nature.  Nevertheless, these projections still provide a useful illustration of how the surrender strategy behaves.

Figure \ref{fig: surrender_rate_bench} illustrates the surrender ratio of simulated paths in each year. In general, the surrender ratio increases with time. However, the curve is not smooth, which is natural in our setting. In the non-Markovian model, the optimal surrender depends on high-dimensional past state variables. The irregularities and non-smoothness arise when we project the high-dimensional past state variables to the one-dimensional time $t$. Despite these fluctuations, the figure shows an upward trend in the surrender ratio over time, consistent with economic intuition: early in the contract period, the embedded guarantees have greater time value, and surrender penalties discourage early surrender. As time progresses, the guarantees gradually lose their time value, and surrender penalties decline, so the policyholder has stronger incentives to surrender the policy to realize the equity return and avoid potential future fees charged by the insurer.


\clearpage
\section{Conclusions}
\label{sec: conclusions}
In this paper, we study the optimal surrender of a variable annuity with GMMB and GMDB under Volterra mortality and the rough Heston model. Both models are non-Markovian and require new computational techniques. To address this difficulty, we propose a deep signature LSMC framework that approximates continuation values with neural networks and the path signatures, solves the optimal surrender problem on a discretized time grid, and finally computes the corresponding fair fee via the bisection method. To ensure the algorithm's stability, we prove the convergence of the deep signature LSMC method via an inductive argument on the difference between the true VA price and the candidate deep-learning VA price. Our numerical analysis shows that the fair fee increases with $H_S$ and $H_m$. Overall, our paper first provides a reliable framework for computing the VA surrender under the non-Markovian processes. It can serve as a benchmark method to further inspire the design of actuarial products under the non-Markovian settings.

\clearpage
\bibliographystyle{apalike}
\bibliography{ref}

\end{document}